\documentclass[11pt]{article}
\usepackage{amsfonts}
\usepackage{amsmath}
\usepackage{geometry}
\usepackage{graphicx}
\usepackage{rotating}

\newtheorem{theorem}{Theorem}

\newtheorem{lemma}{Lemma}

\newtheorem{proposition}{Proposition}
\newtheorem{remark}{Remark}

\newenvironment{proof}[1][Proof]{\noindent\textbf{#1.} }{\ \rule{0.5em}{0.5em}}
\input{tcilatex}
\begin{document}

\title{Integrability and Identification in Multinomial Choice Models}
\author{Debopam Bhattacharya\thanks{%
The author acknowledges financial support from the European Research Council
via a Consolidator Grant EDWEL, Project number 681565.} \\
University of Cambridge}
\date{May 5, 2021}
\maketitle

\begin{abstract}
McFadden's random-utility model of multinomial choice has long been the
workhorse of applied research. We establish shape-restrictions under which
multinomial choice-probability functions can be rationalized via
random-utility models with nonparametric unobserved heterogeneity and
general income-effects. When combined with an additional restriction, the
above conditions are equivalent to the canonical Additive Random Utility
Model. The sufficiency-proof is constructive, and facilitates nonparametric
identification of preference-distributions without requiring
identification-at-infinity type arguments. A corollary shows that
Slutsky-symmetry, a key condition for previous rationalizability results, is
equivalent to absence of income-effects. Our results imply theory-consistent
nonparametric bounds for choice-probabilities on counterfactual budget-sets.
They also apply to widely used random-coefficient models, upon conditioning
on observable choice characteristics. The theory of partial differential
equations plays a key role in our analysis.


JEL\ Codes: C14, C25, D11.
\end{abstract}

\section{Introduction}

The random utility model of multinomial choice (McFadden, 1973) has gained
enormous popularity among applied economists. However, there has been
limited research on the micro-theoretic underpinning of such models, and in
particular, on the question of `integrability', i.e. which choice
probability functions are logically consistent with a random utility model.
Apart from obvious theoretical interest, this question has practical
implications for empirical modelling of individual demand as well as
predicting aggregate demand and welfare on counterfactual budget-sets that
arise from a new tax or subsidy or changes in choice-sets due to addition or
elimination of choice-options. In particular, any utility distribution that
rationalizes a given demand dataset can be used, in addition to shape
restrictions implied by economic theory, to construct nonparametric,
theory-consistent bounds on such counterfactuals.

There has been comparatively more work on integrability in empirical demand
models with \textit{continuous} goods, c.f. Lewbel, 2001. More recently,
Dette, Hoderlein and Neumayer 2016 and Hausman and Newey 2016 have derived
integrability conditions for choice of a single continuous good and
Bhattacharya 2020 has obtained them for binary choice settings under general
(i.e. not necessarily additive) heterogeneity. The multinomial discrete
choice case differs fundamentally from the single continuous good setting
because the price of different alternatives are generically distinct, unlike
continuous choice where the per unit price is constant across choices.

In the present paper, we first show that in multinomial choice settings that
allow for nonparametric unobserved heterogeneity and income effects, there
is a set of shape restrictions on conditional choice probability functions
which together are sufficient for integrability. The proof of this result is
constructive, and the rationalizing utility functions are obtained by
inverting solutions of certain partial differential equations (PDEs). The
way in which PDEs arise here is unrelated to Roy's Identity (c.f. Mas-Colell
et al, 1995, Proposition 3.G.4); the partial derivatives appearing in the
PDE are of the average \textit{demand} function, not the indirect utility
function. Together with an additional restriction, the above conditions are
then shown to be both necessary and sufficient for the canonical additive
random utility model (ARUM) of McFadden. In our analysis of integrability,
we leave the joint distribution of unobserved heterogeneity terms
nonparametric. Unlike the computationally intensive algorithmic approach of
McFadden and Richter 1990, further investigated in Kitamura and Stoye 2018,
our conditions are closed-form and analytic, and can therefore be imposed on
choice probability functions during estimation; they are also global, in the
sense that their forms do not depend on how many and which budget sets
happen to be observed in a specific dataset. On the other hand, MR and KS's
approach work under unrestricted heterogeneity, whereas our set-up is the
canonical model with additive heterogeneity but also covers more flexible
models like the widely-used random coefficient setting (e.g. mixed logit)
which, conditional on observed covariates, have an additive structure.

For discrete choice, Daly and Zachary 1978 provided a set of closed-form,
global conditions under which closed-form choice-probability functions can
be justified as having arisen from preference maximization by a
heterogeneous population. These conditions were independently derived in
Armstrong and Vickers, 2015, who improved upon the Daly-Zachary results by
including an outside option in the choice set. In all of these results, a
key condition for integrability is Slutsky symmetry, analogous to the
classic textbook case for demand systems with continuous goods.\footnote{%
This is distinct from Slutsky \textbf{negativity} c.f. Bhattacharya 2021 for
the general (i.e. not necessarily additive) heterogeneity case. Dagsvik and
Karlstrom 2005 provide some related results for the setting where unobserved
heterogeneity is both additive and is assumed to have \textit{known}
distribution. See also Fosgerau et al 2013 and Delle Site 2014.} As a
corollary of our main theorem, we show that in the multinomial setting,
Daly-Zachary's Slutsky symmetry is equivalent to the absence of income
effects, i.e. that conditional choice probabilities do not depend on the
decision-makers' income. The \textquotedblleft necessity\textquotedblright\
part is easy to show, but showing \textquotedblleft
sufficiency\textquotedblright , i.e. that Slutsky symmetry implies absence
of income effects is non-trivial.

Next, we show how our integrability results can be used to nonparametrically
identify the underlying preference distributions from empirical
choice-probabilities. A key restriction delivering this identification
result -- viz. invertibility of sub-utilities in the numeraire due to
non-satiation -- is based on \textit{economic} theory, as opposed to
statistical assumptions. This is in contrast to existing results on
identification of multinomial choice models, which either rely on
statistical/mathematical assumptions, e.g. utilities being linearly
separable in a covariate with large support, c.f. Matzkin 1993 (see also
Allen and Rehbeck 2019 for related results). An important distinguishing
feature of our set-up is that the arguments of choice-probability functions,
viz. price and income, arise from budget constraints and they play very
specific roles in the proof of integrability and the identification
strategy. In that sense, our approach utilizes the basic economic theory of
utility maximization subject to budget constraints, in contrast to the
approach of Matzkin or Allen and Rehbeck that treat the arguments of
choice-probabilities in a more abstract, statistical way. An important
empirical consequence of this is that our results lead to nonparametric,
theory-consistent bounds for choice probabilities on counterfactual budget
sets. No such bounds on counterfactuals are possible in the set-up of
Matzkin or Allen and Rehbeck unless utility indices and the heterogeneity
distribution are assumed to have a known parametric form. Furthermore, from
a purely methodological standpoint, achieving nonparametric identification
by solving PDEs appears to be novel in the discrete choice literature.

Next, we discuss the empirical usefulness of our results by showing how they
can be used (a) to analyze random coefficient models that are popular in
applied work, e.g. McFadden-Train's mixed logit or the BLP model, and (b) to
calculate theory-consistent bounds for demand and welfare on counterfactual
budget sets, e.g. those resulting from prospective introduction of new taxes
and subsidies, price-changes due to mergers and potential changes in choice
sets e.g. due to removal of alternatives.\smallskip

The plan for the rest of the paper is as follows. Section 2 discusses
integrability for multinomial choice in presence of income effects, and
presents Lemma 1 and Theorem 1, the two key results of this paper, followed
by a discussion of Daly-Zachary's Slutsky symmetry condition and its
connection with lack of income effects. Section 3 discusses four further
points, viz. the implication of the integrability result for nonparametric
identification of preference distributions, incorporation of covariates into
the analysis, the applicability of these results to random coefficient
models and using these results to calculate bounds on counterfactual choice
probabilities. Section 4 concludes. A short appendix at the end presents two
mathematical results on partial and ordinary differential equations that are
intensively used in this paper, as well as proofs of the two main results.

Throughout the paper, we will assume continuous differentiability of the
choice probability function in prices and income to sufficient orders and,
to avoid repetitions, not include this separately each time among the
conditions for our results.

\section{Set-up and Key Results}

Consider a setting of multinomial choice, where the discrete alternatives
are indexed by $j=0,1,...,J$, individual income is $y$, price of alternative 
$j$ is $p_{j}$; if alternative $0$ refers to the outside option, i.e. not
buying any of the alternatives, then $p_{0}\equiv 0$. Let the utility from
consuming the $j$th alternative and a quantity $z$ of the numeraire be given
by $U\left( j,z\right) $, where $U\left( j,\cdot \right) $ is not
necessarily linear. The consumer's problem is $\max_{j\in \left\{
0,1,...,J\right\} ,z}\left[ U\left( j,z\right) +\varepsilon _{j}\right] $,
subject to the budget constraint $z\leq y-p_{j}$, where $y$ is the
consumer's income, $p_{j}$ is the price of alternative $j$ faced by the
consumer, and $\varepsilon _{j}$ is unobserved heterogeneity in the
consumer's preferences. If $U\left( j,\cdot \right) $ is strictly increasing
(i.e. non-satiation in the numeraire), then we can rewrite the consumer
problem as $\max_{j\in \left\{ 0,1,...,J\right\} }\left[ U\left(
j,a_{j}\right) +\varepsilon _{j}\right] $, where $a_{j}\equiv y-p_{j}$, $%
a_{0}=y$. Denote the \textit{structural} probability of choosing alternative 
$j\in \left\{ 0,...,J\right\} $ at $\mathbf{a\equiv }\left(
a_{0},..,a_{J}\right) $ by $q_{j}\left( \mathbf{a}\right) $. In words, if we
randomly sample individuals from the population, and offer the vector $%
\mathbf{a}$ to each sampled individual, then a fraction $q_{j}\left( \mathbf{%
a}\right) $ will choose alternative $j$, in expectation. It is easy to
incorporate other attributes of the alternatives and characteristics of
consumers in our analysis, and we outline how to that in Section 3. For now,
we suppress other covariates for clarity of exposition. Note that the above
structure covers models for bundles, c.f. Gentzkow 2007. For example, if the
choice set is $\left( \left\{ 0\right\} ,\left\{ 1\right\} ,\left\{
2\right\} ,\left\{ 1,2\right\} \right) $, then that model is equivalent to a
multinomial model with 4 alternatives where the price of option $\left\{
1,2\right\} $ is $p_{1}+p_{2}$.

The key question of this paper is whether utility maximization in the above
setting of multinomial choice that allows for income effects (corresponding
to $U\left( j,\cdot \right) $ being nonlinear) impose any restriction on
choice-probabilities. To answer this question, we first introduce a
condition that we call `Slutsky invariance'.\smallskip

\begin{quote}
(A): For any $\mathbf{a}$, and any pair of alternatives $k\neq l$, the ratio 
$\frac{\partial }{\partial a_{k}}q_{l}\left( \mathbf{a}\right) /\frac{%
\partial }{\partial a_{l}}q_{k}\left( \mathbf{a}\right) $ depends only on $%
a_{k}$ and $a_{l}$.\smallskip
\end{quote}

\textbf{Motivation}: To see where this restriction comes from, consider the
above setting of multinomial choice, and let the utility from consuming the $%
j$th alternative and a quantity $z$ of the numeraire be given by $U\left(
j,z\right) +\varepsilon _{j}$. The $\left\{ \varepsilon _{j}\right\} $,
which represent unobserved heterogeneity in preferences, are allowed to have
any arbitrary and unspecified joint distribution in the population (subject
to the resulting choice probability functions being smooth). If $U\left(
j,\cdot \right) $ is strictly increasing, i.e. preferences are non-satiated
in the numeraire, then we can replace $z=y-p_{j}\equiv a_{j}$, and rewrite
the consumer problem as%
\begin{equation}
\max_{j\in \left\{ 0,1,...,J\right\} }\left[ U\left( j,a_{j}\right)
+\varepsilon _{j}\right] .  \label{n}
\end{equation}%
To allow for income effects, we let $U\left( j,a_{j}\right) \equiv
h_{j}\left( a_{j}\right) $, where $h_{j}\left( \cdot \right) $ are smooth,
possibly nonlinear, strictly increasing, \textit{unspecified} functions of
the $a_{j}$'s. When $h_{j}\left( \cdot \right) $ are nonlinear, the
conditional choice-probabilities will depend on income, i.e., there are
non-zero income effects. This structure is also observationally equivalent
to a utility structure where unobserved heterogeneity is not additively
separable from the $a_{j}$'s (see below) in the utility function.

Now, for the above set-up, the choice probability for the $0$th alternative
is given by%
\begin{eqnarray}
&&q_{0}\left( \mathbf{a}\right)  \notag \\
&=&\Pr \left( \cap _{j\neq 0}\left\{ h_{0}\left( a_{0}\right) +\varepsilon
_{0}>h_{j}\left( a_{j}\right) +\varepsilon _{j}\right\} \right)  \notag \\
&=&\Pr \left[ \cap _{j\neq 0}\left\{ h_{0}\left( a_{0}\right) -h_{j}\left(
a_{j}\right) >\varepsilon _{j}-\varepsilon _{0}\right\} \right]  \notag \\
&=&\int_{-\infty }^{\infty }\int_{-\infty }^{\left( h_{0}\left( a_{0}\right)
-h_{1}\left( a_{1}\right) \right) +\varepsilon _{0}}...\int_{-\infty
}^{\left( h_{0}\left( a_{0}\right) -h_{J}\left( a_{J}\right) \right)
+\varepsilon _{0}}g\left( \varepsilon \right) d\varepsilon
_{J}...d\varepsilon _{1}d\varepsilon _{0}\text{.}  \label{d}
\end{eqnarray}%
Therefore, by the first fundamental theorem of calculus,%
\begin{eqnarray}
&&\frac{\partial }{\partial a_{1}}q_{0}\left( \mathbf{a}\right)  \notag \\
&=&-h_{1}^{\prime }\left( a_{1}\right) \left[ 
\begin{array}{c}
\int_{-\infty }^{\infty }\int_{-\infty }^{\substack{ \varepsilon _{0}  \\ %
+h_{0}\left( a_{0}\right)  \\ -h_{2}\left( a_{2}\right) }}...\int_{-\infty } 
^{\substack{ \varepsilon _{0}  \\ +h_{0}\left( a_{0}\right)  \\ -h_{J}\left(
a_{J}\right) }}g\left( 
\begin{array}{c}
\varepsilon _{0}, \\ 
\left( h_{0}\left( a_{0}\right) -h_{1}\left( a_{1}\right) \right)
+\varepsilon _{0}, \\ 
\varepsilon _{2},...\varepsilon _{J}%
\end{array}%
\right) \\ 
d\varepsilon _{J}...d\varepsilon _{2}d\varepsilon _{0}%
\end{array}%
\right] \text{.}  \label{15}
\end{eqnarray}%
Similarly,%
\begin{equation*}
q_{1}\left( \mathbf{a}\right) =\int_{-\infty }^{\infty }\int_{-\infty } 
^{\substack{ \varepsilon _{1}  \\ +h_{1}\left( a_{1}\right)  \\ -h_{0}\left(
a_{0}\right) }}...\int_{-\infty }^{\substack{ \varepsilon _{1}  \\ %
+h_{1}\left( a_{1}\right)  \\ -h_{J}\left( a_{J}\right) }}g\left(
\varepsilon \right) d\varepsilon _{J}...d\varepsilon _{2}d\varepsilon
_{0}d\varepsilon _{1}\text{,}
\end{equation*}%
implying by the first fundamental theorem and chain-rule that%
\begin{eqnarray}
&&\frac{\partial }{\partial a_{0}}q_{1}\left( \mathbf{a}\right)  \notag \\
&=&-h_{0}^{\prime }\left( a_{0}\right) \int_{-\infty }^{\infty
}\int_{-\infty }^{\substack{ h_{1}\left( a_{1}\right)  \\ -h_{2}\left(
a_{2}\right) +\varepsilon _{1}}}...\int_{-\infty }^{\substack{ h_{1}\left(
a_{1}\right)  \\ -h_{J}\left( a_{J}\right) +\varepsilon _{1}}}g\left( 
\begin{array}{c}
h_{1}\left( a_{1}\right) -h_{0}\left( a_{0}\right) +\varepsilon _{1}, \\ 
\varepsilon _{1},\varepsilon _{2},...\varepsilon _{J}%
\end{array}%
\right) d\varepsilon _{J}...d\varepsilon _{2}d\varepsilon _{1}  \label{a} \\
&&\overset{(1)}{=}-h_{0}^{\prime }\left( a_{0}\right) 
\begin{array}{c}
\int_{-\infty }^{\infty }\int_{-\infty }^{\substack{ s_{0}+h_{0}\left(
a_{0}\right)  \\ -h_{2}\left( a_{2}\right) }}...\int_{-\infty
}^{s_{0}+h_{0}\left( a_{0}\right) -h_{J}\left( a_{J}\right) }g\left( 
\begin{array}{c}
s_{0}, \\ 
s_{0}-h_{1}\left( a_{1}\right) +h_{0}\left( a_{0}\right) , \\ 
\varepsilon _{2},...\varepsilon _{J}%
\end{array}%
\right) \\ 
d\varepsilon _{J}...d\varepsilon _{2}ds_{0}%
\end{array}
\notag \\
&=&\frac{h_{0}^{\prime }\left( a_{0}\right) }{h_{1}^{\prime }\left(
a_{1}\right) }\frac{\partial }{\partial a_{1}}q_{0}\left( \mathbf{a}\right) 
\text{, using (\ref{15}),}  \notag
\end{eqnarray}%
where the second equality $\overset{(1)}{=}$ follows by substituting $%
s_{0}=h_{1}\left( a_{1}\right) -h_{0}\left( a_{0}\right) +\varepsilon _{1}$
in (\ref{a}).

The same argument can be repeated for any other pair of alternatives $l\neq
k $, to obtain%
\begin{equation}
\frac{\frac{\partial }{\partial a_{k}}q_{l}\left( \mathbf{a}\right) }{\frac{%
\partial }{\partial a_{l}}q_{k}\left( \mathbf{a}\right) }=\frac{%
h_{k}^{\prime }\left( a_{k}\right) }{h_{l}^{\prime }\left( a_{l}\right) }%
\text{,}  \label{b}
\end{equation}%
for all $\mathbf{a}$, and it is clear that the RHS\ of (\ref{b}) depends
only on $a_{k}$ and $a_{l}$, and thus satisfies condition (A)
above.\smallskip

\begin{remark}
Condition (\ref{b}) has no relation with the Independence of Irrelevant
Alternatives (IIA) property. Indeed, the model above will \textbf{not} have
the IIA property if the $\varepsilon _{j}$s are correlated across
alternatives (i.e. across $j$), but it will continue to satisfy (\ref{b}),
since uncorrelatedness of $\varepsilon $s was not used to derive (\ref{b}%
).\smallskip
\end{remark}

\textbf{Main Results}: We now state and prove our main results. The first
result is that the Slutsky invariance condition stated above, plus two
shape-restrictions on $q_{j}\left( \mathbf{\cdot }\right) $'s are jointly
sufficient for integrability, i.e., under those restrictions on $q_{j}\left( 
\mathbf{\cdot }\right) $'s, we can find a set of utility functions and a
joint distribution of unobserved preference heterogeneity, such that
individual maximization of these utilities will indeed produce the
conditional choice-probabilities $\left\{ q_{j}\left( \mathbf{\cdot }\right)
\right\} $, $j=0,1,...,J$.\smallskip

To state and prove our first result, we will use the following additional
notation: let $\mathbf{a}_{-j}$ denote the vector $\left(
a_{0},a_{1},...a_{j-1},a_{j+1},...a_{J}\right) $ and let for each $%
j=0,1,...J $, $\lim_{\mathbf{a}_{-j}\downarrow \mathbf{c}^{(j)}\left(
a_{j}\right) }$ denote that each $k$th component of $\mathbf{a}_{-j}$ goes
to a constant $c_{k}^{\left( j\right) }\left( a_{j}\right) $ with $\mathbf{c}%
^{(j)}\left( a_{j}\right) =\left( c_{0}^{\left( j\right) }\left(
a_{j}\right) ,...,c_{j-1}^{\left( j\right) }\left( a_{j}\right)
,c_{j+1}^{\left( j\right) }\left( a_{j}\right) ,...,c_{J}^{\left( j\right)
}\left( a_{j}\right) \right) $. Similarly, $\lim_{a_{j}\downarrow d^{\left(
j\right) }\left( \mathbf{a}_{-j}\right) }$ denotes that for fixed $\mathbf{a}%
_{-j}$, $a_{j}$ decreases to a constant $d^{\left( j\right) }\left( \mathbf{a%
}_{-j}\right) $ (whose value depends on $\mathbf{a}_{-j}$).

\begin{lemma}
Suppose that the following three conditions are satisfied by the
choice-probabilities $\left\{ q_{j}\left( \mathbf{a}\right) \right\} $:

(i) For each $j=0,1,...,J$, and each $\mathbf{a}$, $q_{j}\left( \mathbf{a}%
\right) $ is strictly increasing in $a_{j}$ and strictly decreasing in $%
a_{k} $ for $k\neq j$, continuously differentiable in each argument, and for
all $j $, there exist a vector of constants $\mathbf{c}^{(j)}\left(
a_{j}\right) $ and a constant $d^{\left( j\right) }\left( \mathbf{a}%
_{-j}\right) $ in the supports of $\left\{ Y-P_{j}\right\} $ such that $%
\lim_{\mathbf{a}_{-j}\downarrow \mathbf{c}^{(j)}\left( a_{j}\right)
}q_{j}\left( \mathbf{a}\right) =1$ and $\lim_{a_{j}\downarrow d^{\left(
j\right) }\left( \mathbf{a}_{-j}\right) }q_{j}\left( \mathbf{a}\right)
=0=1-\lim_{a_{j}\uparrow \infty }q_{j}\left( \mathbf{a}\right) $;

(ii) For any pair of alternatives $j\neq m$ and any $\mathbf{a}$ satisfying $%
\frac{\partial }{\partial a_{j}}q_{m}\left( \mathbf{a}\right) \neq 0$, the
ratio $\frac{\partial }{\partial a_{m}}q_{j}\left( \mathbf{a}\right) /\frac{%
\partial }{\partial a_{j}}q_{m}\left( \mathbf{a}\right) $ does not depend on 
$a_{k}$, for $k\notin \left\{ m,j\right\} $, and has uniformly bounded
derivatives with respect to $a_{m}$ and $a_{j}$;

(iii) for each $r=0,1,...J$, the $J$th order cross partial derivatives $%
\frac{\partial ^{J}}{\partial a_{0}\partial a_{1}...\partial a_{r-1}\partial
a_{r+1}...\partial a_{J}}q_{r}\left( \mathbf{a}\right) $ exist, are
continuous, and satisfy $\left( -1\right) ^{J}\frac{\partial ^{J}}{\partial
a_{0}\partial a_{1}...\partial a_{r-1}\partial a_{r+1}...\partial a_{J}}%
q_{r}\left( \mathbf{a}\right) \geq 0$.

Then there exist random variables $\mathbf{V}=\left(
V_{0},V_{1},...,V_{m-1},V_{m+1},...,V_{J}\right) $ with support $\mathcal{V}%
\sqsubseteq $ $\mathbb{R}^{J}$ and a joint density function $f\left( \cdot
\right) $, and `utility' functions $w_{j}\left( a,v_{j}\right) :\mathbb{R}%
\times \mathcal{V}_{j}\rightarrow \mathbb{R}$, such that $w_{j}\left( \cdot
,v_{j}\right) $ are strictly increasing and continuous, $w_{m}\left(
a_{m},v_{m}\right) \equiv a_{m}$, and 
\begin{equation*}
q_{j}\left( a_{0},a_{1},...,a_{J}\right) =\int_{\mathcal{V}}\cap _{k\neq
j}1\left\{ w_{j}\left( a_{j},v_{j}\right) \geq w_{k}\left(
a_{k},v_{k}\right) \right\} f\left( \mathbf{v}\right) d\mathbf{v}
\end{equation*}%
for each $j=0,1,...J$. Thus the utility functions $\left\{ w_{j}\left(
a,v_{j}\right) \right\} $ and heterogeneity distribution $f\left( \cdot
\right) $ rationalize the choice probabilities $\left\{ q_{j}\left( \mathbf{a%
}\right) \right\} $. (\textbf{Proof in Appendix})\smallskip
\end{lemma}

Condition (i) is intuitive, and corresponds to preferences being
non-satiated in the quantity of numeraire. Indeed, if choice probabilities
are generated by the structure%
\begin{equation*}
q_{j}\left( \mathbf{a}\right) =\int_{\mathcal{V}}1\left\{ W_{j}\left(
a_{j},\eta \right) \geq \max_{r\in \left\{ 0,1,...J\right\} \backslash
\left\{ j\right\} }W_{r}\left( a_{r},\eta \right) \right\} f\left( \eta
\right) d\eta \text{,}
\end{equation*}%
where $W_{j}\left( ,\eta \right) $ are strictly increasing and continuous,
and their distributions sufficiently smooth, then condition (i) must hold.
The limiting condition $\lim_{\mathbf{a}_{-j}\downarrow \mathbf{c}%
^{(j)}\left( a_{j}\right) }q_{j}\left( \mathbf{a}\right) =1$ means that
holding $a_{j}$ fixed, if we lower $\left\{ a_{k},k\neq j\right\} $
sufficiently, then the probability of choosing $j$ rises to 1. For example,
if the price of each alternative $k\neq j$ becomes sufficiently high, then
eventually everyone will choose $j$. Similarly, $\lim_{a_{j}\downarrow
d^{\left( j\right) }\left( \mathbf{a}_{-j}\right) }q_{j}\left( \mathbf{a}%
\right) =0$ means that holding income and prices of other alternatives
fixed, if the price of the $j$th alternative increases sufficiently, then
its aggregate demand will become zero. Condition (iii) is related to the
existence of a density function for unobserved heterogeneity. For models
with \textit{parametrically specified }heterogeneity distributions,
condition (iii) was previously used to recover underlying utility functions
(c.f. McFadden, 1978 just above Eqn. 12, and McFadden 1981). The motivation
for condition (ii) was discussed right before Lemma 1. The proof of this
lemma, detailed in the appendix, is based on differentiating the identity $%
\sum_{j=0}^{J}q_{j}\left( \mathbf{a}\right) =1$, applying condition (ii) and
solving the resulting partial differential equation.

Note that by using the utility functions and heterogeneity distribution
obtained via Lemma 1, one can simulate choice probabilities at the observed $%
\mathbf{a}$'s. To do this, for any pair of alternatives $j\neq m$ a least
squares projection of $\frac{\partial }{\partial a_{m}}q_{j}\left( \mathbf{a}%
\right) /\frac{\partial }{\partial a_{j}}q_{m}\left( \mathbf{a}\right) $ on
a polynomial sieve in $a_{j},a_{m}$ would be used to generate the
coefficient functions of the PDEs, which are then solved to obtain the
utility functions and the heterogeneity distribution (see the section
"Identification" below for further details), as in Lemma 1. One can then
test whether these simulated choice probabilities equal the observed
choice-probabilities. Passing this test would imply that the observed choice
probabilities can be rationalized.\smallskip

\begin{remark}
The \textbf{utility}\textit{\ }function for each alternative $j$, viz. $%
w_{j}\left( a_{j},v_{j}\right) $, constructed in the proof of Lemma 1,
consists of a scalar heterogeneity $v_{j}$. However, the individual \textbf{%
demand}\textit{\ }function for alternative $j$ has $J$ separate sources of
heterogeneity, i.e.%
\begin{eqnarray*}
Q_{j}\left( \mathbf{a,v}\right) &=&1\left\{ w_{j}\left( a_{j},v_{j}\right)
\geq \max_{r\in \left\{ 0,1,...J\right\} \backslash \left\{ j\right\}
}w_{r}\left( a_{r},v_{r}\right) \right\} \\
&=&Q_{j}\left( a_{0},a_{1},...a_{J},\underset{J\text{ dimensional
heterogeneity}}{\underbrace{v_{1},v_{2},...,v_{J}}}\right)
\end{eqnarray*}%
Thus, we have rationalized a $\left( J+1\right) $ dimensional choice
probability function via a $J$-dimensional heterogeneity
distribution.\medskip
\end{remark}

The above result establishes a set of conditions for a choice probability
function to be rationalized via a random utility model. The constructed
model, however, is not linear in unobserved heterogeneity. The next result
shows that when combined with an additional requirement, the three
conditions above are necessary and sufficient for integrability via an
additive random utility model.

\begin{theorem}
Assume the same set-up as in Lemma 1, and assume that Conditions (i) and
(iii) of Lemma 1 hold. Additionally it holds that for all $j\neq m$, (ii') $%
\frac{\partial }{\partial a_{m}}q_{j}\left( \mathbf{a}\right) /\frac{%
\partial }{\partial a_{j}}q_{m}\left( \mathbf{a}\right) $ depends only on $%
a_{j},a_{m}$, and is of the form 
\begin{equation*}
\frac{\partial }{\partial a_{m}}q_{j}\left( \mathbf{a}\right) /\frac{%
\partial }{\partial a_{j}}q_{m}\left( \mathbf{a}\right) =G_{m}\left(
a_{m}\right) /G_{j}\left( a_{j}\right) \text{,}
\end{equation*}%
where $G_{j}\left( \cdot \right) ,G_{0}\left( \cdot \right) >0$, for all $%
j\neq m$.\footnote{%
Equivalently,%
\begin{equation*}
\frac{\partial }{\partial a_{m}}q_{j}\left( \mathbf{a}\right) /\frac{%
\partial }{\partial a_{j}}q_{m}\left( \mathbf{a}\right) =H_{m}\left(
a_{m}\right) \times H_{j}\left( a_{j}\right) \text{,}
\end{equation*}%
or equivalently,%
\begin{equation*}
\ln \left( \frac{\partial }{\partial a_{m}}q_{j}\left( \mathbf{a}\right) /%
\frac{\partial }{\partial a_{j}}q_{m}\left( \mathbf{a}\right) \right)
=h_{m}\left( a_{m}\right) +h_{j}\left( a_{j}\right) \text{,}
\end{equation*}%
where $H_{m}\left( \cdot \right) $ and $H_{j}\left( \cdot \right) $ are
positive functions, and $h_{m}\left( \cdot \right) $, $h_{j}\left( \cdot
\right) $ are real-valued functions.} Then there exist strictly increasing
utility functions $U\left( j,\cdot \right) :\mathbb{R}\rightarrow \mathbb{R}$%
, and $J$ dimensional unobserved heterogeneity $\left(
v_{1},...,v_{J}\right) \equiv \left( \varepsilon _{1}-\varepsilon
_{0},...,\varepsilon _{J}-\varepsilon _{0}\right) $ with continuous density
such that for all $j=0,1,...,J$.%
\begin{equation}
q_{j}\left( a_{0},a_{1},...,a_{J}\right) =\Pr \left[ \cap _{k\neq j}1\left\{
U\left( j,a_{j}\right) +\varepsilon _{j}\geq U\left( k,a_{k}\right)
+\varepsilon _{k}\right\} \right] \text{.}  \label{2}
\end{equation}%
Conditions (i), (ii'), (iii) are also necessary for (\ref{2}) to hold (proof
in Appendix).\smallskip
\end{theorem}

\textbf{Conditions in standard form}: We have expressed choice probabilities
as functions of the $a_{j}$s, as opposed to $p_{j}$s and $y$, since it is
more natural to impose monotonicity of \ a function in its arguments, rather
than on combination of derivatives with respect to arguments. If choice
probabilities are instead expressed in the standard form with income and
prices as arguments, one has%
\begin{eqnarray*}
q_{j}\left( a_{0},a_{1},...,a_{J}\right) &=&\bar{q}_{j}\left(
a_{0},a_{0}-a_{1},...,a_{0}-a_{J}\right) \\
&=&\bar{q}_{j}\left( y,p_{1},...,p_{J}\right) \equiv \bar{q}_{j}\left( y,%
\mathbf{p}\right) \text{.}
\end{eqnarray*}%
Then the shape restrictions, i.e. condition (i) become: for each $j=1,...J$, 
$\partial \bar{q}_{j}\left( y,\mathbf{p}\right) /\partial p_{j}\leq 0$, $%
\partial \bar{q}_{j}\left( \mathbf{p},y\right) /\partial p_{k}\geq 0$ for
all $k\neq j$, and $\sum_{k=1}^{J}\partial \bar{q}_{j}\left( y,\mathbf{p}%
\right) /\partial p_{k}+\partial \bar{q}_{j}\left( y,\mathbf{p}\right)
/\partial y\leq 0$ for all $j=1,...J$. The forms of these expressions bear
similarity to Slutsky inequality conditions in standard, deterministic
demand analysis for continuous goods. An important difference with the
standard continuous case is that our condition is%
\begin{equation}
\sum_{k=1}^{J}\partial \bar{q}_{j}\left( y,\mathbf{p}\right) /\partial
p_{k}+\partial \bar{q}_{j}\left( y,\mathbf{p}\right) /\partial y\leq 0\text{,%
}
\end{equation}%
in contrast to the standard continuous case where the Slutsky condition is%
\begin{equation}
\sum_{k=1}^{J}\partial \bar{q}_{j}\left( y,\mathbf{p}\right) /\partial p_{k}+%
\bar{q}_{j}\left( y,\mathbf{p}\right) \times \partial \bar{q}_{j}\left( y,%
\mathbf{p}\right) /\partial y\leq 0.
\end{equation}

Condition (ii) becomes: for all $j=1,2,...,J$,%
\begin{equation*}
\frac{\sum_{k=1}^{J}\partial \bar{q}_{j}\left( y,\mathbf{p}\right) /\partial
p_{k}+\partial \bar{q}_{j}\left( y,\mathbf{p}\right) /\partial y}{\partial 
\bar{q}_{0}\left( y,\mathbf{p}\right) /\partial p_{j}}
\end{equation*}%
depends on $\left( y,\mathbf{p}\right) $ only via $\left( y,y-p_{j}\right) $%
, i.e. via $\left( y,p_{j}\right) $, and for all $j,k=1,2,...,J$ with $j\neq
k$, $\frac{\partial \bar{q}_{j}\left( y,\mathbf{p}\right) /\partial p_{k}}{%
\partial \bar{q}_{k}\left( y,\mathbf{p}\right) /\partial p_{j}}$ depends on $%
\left( y,\mathbf{p}\right) $ only via $\left( y-p_{k},y-p_{j}\right) $.
Condition (iii') strengthens to $\frac{\sum_{k=1}^{J}\partial \bar{q}%
_{j}\left( y,\mathbf{p}\right) /\partial p_{k}+\partial \bar{q}_{j}\left( y,%
\mathbf{p}\right) /\partial y}{\partial \bar{q}_{0}\left( y,\mathbf{p}%
\right) /\partial p_{j}}$ being of the form $h_{0}\left( y\right) \times
h_{j}\left( y-p_{j}\right) $ for each $j=1,...,J$ and $\frac{\partial \bar{q}%
_{j}\left( y,\mathbf{p}\right) /\partial p_{k}}{\partial \bar{q}_{k}\left( y,%
\mathbf{p}\right) /\partial p_{j}}$ is of the form $h_{k}\left(
y-p_{k}\right) \times h_{j}\left( y-p_{j}\right) $ for all $j,k=1,2,...,J$
with $j\neq k$. Finally, condition (iii) is: for all $r=1,2,...,J$,%
\begin{equation*}
\sum_{k=1}^{J}\frac{\partial }{\partial p_{k}}\left[ \frac{\partial ^{J-1}}{%
\partial p_{1}...\partial p_{r-1}\partial p_{r+1}...\partial p_{J}}\bar{q}%
_{r}\left( y,\mathbf{p}\right) \right] \geq 0.
\end{equation*}

\subsection{Daly-Zachary's Slutsky-Symmetry}

In the above set-up, Daly-Zachary's Slutsky symmetry conditions are that for
any two alternatives $k,l\in \left\{ 0,1,...,J\right\} $, $k\neq l$,%
\begin{equation}
\frac{\partial }{\partial a_{l}}q_{k}\left( \mathbf{a}\right) =\frac{%
\partial }{\partial a_{k}}q_{l}\left( \mathbf{a}\right) \text{.\footnotemark 
}  \label{S}
\end{equation}%
\footnotetext{%
Daly-Zachary defines choice probabilities as functions of price and income, $%
\bar{q}_{j}\left( p_{0},p_{1},...,p_{J},y\right) $. This is equivalent to
our notation of $q_{j}\left( a_{0},a_{1},...a_{J}\right) $ with $a_{0}=y$, $%
a_{1}=y-p_{1}$,...,$a_{J}=y-p_{J}$, in that one can move back and forth
between the two notations, since%
\begin{eqnarray*}
q_{j}\left( a_{0},a_{1},...,a_{J}\right) &=&\bar{q}_{j}\left(
a_{0}-a_{1},a_{0}-a_{2},...,a_{0}-a_{J}\right) \text{, and} \\
\bar{q}_{j}\left( p_{1},p_{2},...,p_{J},y\right) &=&q_{j}\left(
y,y-p_{1},y-p_{2},...,y-p_{J}\right) \text{.}
\end{eqnarray*}%
\textquotedblleft Slutsky symmetry\textquotedblright\ in Daly-Zachary's
notation is that $\partial \bar{q}_{k}/\partial p_{j}=\partial \bar{q}%
_{j}/\partial p_{k}$ for all $j\neq k$ (if alternative $0$ is the ouside
option, then the corresponding condition is $\partial \bar{q}_{0}/\partial
p_{j}=\partial \bar{q}_{j}/\partial y$). which is identical to (\ref{S}) in
our notation.}We first show that the classic random utility model with no
income effects implies (\ref{S}). We then show that Slutsky symmetry (\ref{S}%
) implies absence of income effects.\smallskip

\textbf{Necessity}: The canonical random utility model of multinomial choice
assumes that the systematic part of the utility from consuming the $j$th
alternative at income $y$ and price $p_{j}$ is given by%
\begin{equation}
U\left( j,a_{j}\right) \equiv a_{j}\text{,}  \label{N}
\end{equation}%
where $a_{j}=y-p_{j}$ as above. Income effects are zero since demand depends
on the $a$'s via the differences $a_{j}-a_{k}=\left( y-p_{j}\right) -\left(
y-p_{k}\right) =p_{k}-p_{j}$. Then (\ref{b}) with $h_{j}\left( a_{j}\right)
=a_{j}$, i.e. $h_{j}^{\prime }\left( a_{j}\right) =1$ implies 
\begin{equation}
\frac{\frac{\partial }{\partial a_{k}}q_{l}\left( \mathbf{a}\right) }{\frac{%
\partial }{\partial a_{l}}q_{k}\left( \mathbf{a}\right) }=1\text{,}
\label{I}
\end{equation}%
for all $\mathbf{a}$. This shows that in the canonical random utility model
with no income effects, Daly-Zachary's Slutsky symmetry condition
holds.\smallskip

\begin{proposition}
\textbf{\ (Sufficiency)}: In the above set-up, Daly-Zachary's Slutsky
symmetry implies absence of income effects.
\end{proposition}

\begin{proof}
First note that because $\sum_{k=0}^{J}q_{k}\left( \mathbf{a}\right) =1$,
differentiating both sides w.r.t. $a_{l}$ gives%
\begin{equation}
\frac{\partial }{\partial a_{l}}q_{l}\left( \mathbf{a}\right)
+\sum_{k=0,k\neq l}^{J}\frac{\partial }{\partial a_{l}}q_{k}\left( \mathbf{a}%
\right) =0\text{.}  \label{12}
\end{equation}%
Substituting (\ref{S}) in (\ref{12}), we get: 
\begin{equation}
\frac{\partial }{\partial a_{l}}q_{l}\left( \mathbf{a}\right)
+\sum_{k=0,k\neq l}^{J}\frac{\partial }{\partial a_{k}}q_{l}\left( \mathbf{a}%
\right) =0\text{.}  \label{T}
\end{equation}%
This is a linear, homogeneous partial differential equation in $q_{l}\left(
\cdot \right) $, and can be solved via the method of characteristics (c.f.
Courant, 1962, Chapter I.5 and II.2, summarized briefly in the Appendix).
The characteristic curve, i.e. the $J$-dimensional subspace on which $%
q_{l}\left( \mathbf{a}\right) $ remains constant, can be obtained by solving
the so-called \textquotedblleft characteristic\textquotedblright\ Ordinary
Differential Equations (see appendix):%
\begin{equation}
\frac{da_{k}}{da_{l}}=1\text{, }k=0,...l-1,l+1,...,J\text{,}  \label{C}
\end{equation}%
with generic solutions $a_{k}-a_{l}=c_{k}$, $k=0,...l-1,l+1,...,J$. This
means that general solutions to (\ref{T}) are of the form%
\begin{equation}
q_{l}\left( \mathbf{a}\right) =H^{l}\left(
a_{0}-a_{l},a_{1}-a_{l},...,a_{l-1}-a_{l},a_{l+1}-a_{l},....a_{J}-a_{l}%
\right) \text{,}  \label{W}
\end{equation}%
where $H^{l}\left( \cdot \right) $ is any arbitrary continuously
differentiable function. Thus $q_{l}\left( \mathbf{a}\right) $ depends on
the $\left( J+1\right) $-dimensional argument $\left(
a_{0},a_{1},a_{2},...a_{J}\right) $ through a $J$-dimensional vector%
\begin{equation*}
\left(
a_{0}-a_{l},a_{1}-a_{l},a_{2}-a_{l},...,a_{l-1}-a_{l},a_{l+1}-a_{l},....a_{J}-a_{l}\right) 
\text{.}
\end{equation*}%
That (\ref{W}) is a solution to (\ref{T}) can also be verified directly by
partially differentiating the RHS of (\ref{W}), and verifying that it
satisfies (\ref{T}). Finally, note that%
\begin{eqnarray*}
&&\left(
a_{0}-a_{l},a_{1}-a_{l},...,a_{l-1}-a_{l},a_{l+1}-a_{l},....a_{J}-a_{l}%
\right) \\
&=&\left(
p_{l},p_{l}-p_{1},...,p_{l}-p_{l-1},p_{l}-p_{l+1},....p_{l}-p_{J}\right) 
\text{,}
\end{eqnarray*}%
and so (\ref{W}) implies that $q_{l}\left( \mathbf{a}\right) $ does not
depend on income. Since $l$ is arbitrary, we have shown that Slutsky
symmetry implies that income effects are absent.
\end{proof}

\section{Further Points}

\subsection{Identification\label{Id}}

Lemma 1 can be used to identify utilities and the heterogeneity
distributions nonparametrically from choice-probabilities observed in a
dataset. Nonparametric identification of multinomial choice models (without
any discussion of integrability) has been studied previously in the
econometric literature, c.f. Matzkin, 1993, 2007 and Allen and Rehbeck,
2019. Since our proof of integrability presented in Lemma 1 is constructive,
it provides an alternative and novel way to obtain identification by solving
PDEs. Unlike Matzkin 1993, our identification strategy does not rely on
identification-at-infinity type arguments nor on linear separability in a
regressor with large support (c.f. Matzkin 2007), but does require
smoothness.

Specifically, our identification approach is as follows. Suppose that the
choice-probabilities are generated by maximization of the utilities $%
u_{j}\equiv \left\{ h_{j}\left( a_{j}\right) +\varepsilon _{j}\right\} $, $%
j=0,...,J$, where the utility functions $h_{j}\left( \cdot \right) $ are
strictly increasing and continuous and hence invertible, but otherwise
unknown. Observe that an observationally equivalent utility structure is
where utility for the $0$th alternative is $a_{0}$ and that for the $j$th
alternative is $h_{0}^{-1}\left( h_{j}\left( a_{j}\right) +\underset{v_{j}}{%
\underbrace{\varepsilon _{j}-\varepsilon _{0}}}\right) \equiv w_{j}\left(
a_{j},v_{j}\right) $, in that these utilities will produce exactly the same
choice probabilities as the $\left\{ u_{j}\right\} $s. We work under this
normalization from now on. We also note in passing that the $w_{j}\left(
a_{j},v_{j}\right) $ are not necessarily additive in the unobserved
heterogeneity $v_{j}$.

Let $\mathbf{a}$ and $q_{j}\left( \mathbf{a}\right) $ be as above. We can
use the proof of Lemma 1 to identify the $w_{j}\left( a_{j},v_{j}\right) $
functions and the joint distribution of $\left( v_{1},...,v_{J}\right) $
from the $\left\{ q_{j}\left( \mathbf{a}\right) \right\} $, as follows.
First, note that 
\begin{equation*}
q_{0}\left( \mathbf{a}\right) =\Pr \left( \cap _{j\neq 0}\left\{
a_{0}>w_{j}\left( a_{j},v_{j}\right) \right\} \right) =\Pr \left[ \cap
_{j\neq 0}\left\{ v_{j}<\omega _{j}\left( a_{j},a_{0}\right) \right\} \right]
\text{,}
\end{equation*}%
so that%
\begin{equation}
\frac{\partial }{\partial a_{j}}q_{0}\left( \mathbf{a}\right) =\frac{%
\partial }{\partial a_{j}}\omega _{j}\left( a_{j},a_{0}\right) \times
F_{j}\left( \omega _{1}\left( a_{1},a_{0}\right) ,...,\omega _{J}\left(
a_{J},a_{0}\right) \right) \text{,}  \label{14a}
\end{equation}%
where $F_{j}\left( \cdot \right) $ denotes the derivative of the joint
distribution function of $\mathbf{v}$ w.r.t. its $j$th element. On the other
hand,%
\begin{eqnarray*}
q_{j}\left( \mathbf{a}\right) &=&\Pr [w_{j}\left( a_{j},v_{j}\right)
>a_{0},w_{j}\left( a_{j},v_{j}\right) >w_{1}\left( a_{1},v_{1}\right)
,...w_{j}\left( a_{j},v_{j}\right) >w_{J}\left( a_{J},v_{J}\right) \\
&=&\Pr [v_{j}>\omega _{j}\left( a_{j},a_{0}\right) ,v_{1}<\omega _{1}\left(
a_{1},w_{j}\left( a_{j},v_{j}\right) \right) ,...v_{J}<\omega _{J}\left(
a_{J},w_{j}\left( a_{j},v_{j}\right) \right) \\
&=&\int_{\omega _{j}\left( a_{j},a_{0}\right) }^{\infty }\int_{-\infty
}^{\omega _{1}\left( a_{1},w_{j}\left( a_{j},v_{j}\right) \right)
}...\int_{-\infty }^{\omega _{J}\left( a_{J},w_{j}\left( a_{j},v_{j}\right)
\right) }f\left( v_{1},...,v_{J}\right) dv_{J}...dv_{1}dv_{j}\text{,}
\end{eqnarray*}%
and therefore, by the chain-rule, the first fundamental theorem of calculus,
and using $w_{j}\left( a_{j},\omega _{j}\left( a_{j},a_{0}\right) \right)
=a_{0}$, we have that 
\begin{eqnarray}
\frac{\partial }{\partial a_{0}}q_{j}\left( \mathbf{a}\right) &=&-\frac{%
\partial }{\partial a_{0}}\omega _{j}\left( a_{j},a_{0}\right) \times
\int_{-\infty }^{\omega _{1}\left( a_{1},a_{0}\right) }...\int_{-\infty
}^{\omega _{J}\left( a_{J},a_{0}\right) }f\left( v_{1},...,v_{J}\right)
dv_{J}...dv_{1}  \notag \\
&=&-\frac{\partial }{\partial a_{0}}\omega _{j}\left( a_{j},a_{0}\right)
\times F_{j}\left( \omega _{1}\left( a_{1},a_{0}\right) ,...,\omega
_{J}\left( a_{J},a_{0}\right) \right) \text{,}  \label{14b}
\end{eqnarray}%
and thus from (\ref{14a}) and (\ref{14b}), we have that%
\begin{equation}
-\frac{\partial \omega _{j}\left( a_{j},a_{0}\right) }{\partial a_{0}}/\frac{%
\partial \omega _{j}\left( a_{j},a_{0}\right) }{\partial a_{j}}\equiv \frac{%
\partial }{\partial a_{0}}q_{j}\left( \mathbf{a}\right) /\frac{\partial }{%
\partial a_{j}}q_{0}\left( \mathbf{a}\right) \text{,}  \label{11}
\end{equation}%
which is the same as (\ref{2}). The RHS of (\ref{11}) is nonparametrically
identifiable from the data, and under the hypothesis of the model, is solely
a function of $a_{0}$ and $a_{j}$, which is a testable implication. If this
implication is not rejected, denote the RHS of (\ref{11}) as $t_{j}\left(
a_{j},a_{0}\right) $ (this $t_{j}\left( \cdot ,\cdot \right) $\ can be
estimated by, say a least squares projection of $\frac{\partial }{\partial
a_{0}}q_{j}\left( \mathbf{a}\right) /\frac{\partial }{\partial a_{j}}%
q_{0}\left( \mathbf{a}\right) $ on a polynomial sieve in $a_{j},a_{0}$).
Then solve the PDE%
\begin{equation*}
\frac{\partial \omega _{j}\left( a_{j},a_{0}\right) }{\partial a_{0}}+\frac{%
\partial \omega _{j}\left( a_{j},a_{0}\right) }{\partial a_{j}}t_{j0}\left(
a_{j},a_{0}\right) =0\text{,}
\end{equation*}%
for the $\omega _{j}\left( \cdot ,\cdot \right) $'s as outlined in the proof
of Lemma 1 below (see (\ref{v}) and (\ref{vi})), where $\omega _{j}\left(
a_{j},a_{0}\right) $ is strictly increasing in $a_{0}$ and strictly
decreasing in $a_{j}$, and obtain the $w_{j}\left( a_{j},v_{j}\right) $ by
inverting the solution $\omega _{j}\left( a_{j},a_{0}\right) $'s w.r.t. $%
a_{0}$, and the joint density of $\mathbf{v}$ using (\ref{6}).

\subsection{Incorporating Covariates}

In our discussion above, choice probabilities $q_{j}\left( \cdot \right) $
defined in Section 2, correspond to so-called \textquotedblleft average
structural function\textquotedblright , c.f. Blundell and Powell 2003, 2004.
Estimating these from a non-experimental dataset might be non-trivial when
observed budget sets (i.e. price and/or income) are correlated with
unobserved individual preferences across the cross-section of consumers. A
common empirical assumption is that budget sets and preferences are
independent, conditional on a set of observed covariates. Hence it is useful
to see how to adapt the above results to the presence of covariates.

Suppose in addition to price and income, we also observe a vector of
characteristics $z_{j}$ for each alternative $j=1,...,J$. Assume that the
choice-probabilities are generated by maximization of the utilities%
\begin{equation}
u_{0}\equiv \left\{ h_{0}\left( a_{0}\right) +\varepsilon _{0}\right\} ,%
\text{ \ \ }u_{j}\equiv \left\{ h_{j}\left( a_{j},z_{j}\right) +\varepsilon
_{j}\right\} ,j=1,...,J,  \label{18}
\end{equation}%
where $h_{0}\left( a\right) $ and each $h_{j}\left( a,z\right) $ are
strictly increasing and continuous in $a$, and hence invertible. Then an
observationally equivalent utility structure is where utility for the $0$th
alternative is $a_{0}$ and that for the $j$th alternative is%
\begin{equation}
h_{0}^{-1}\left( h_{j}\left( a_{j},z_{j}\right) +\underset{v_{j}}{%
\underbrace{\varepsilon _{j}-\varepsilon _{0}}}\right) \equiv w_{j}\left(
a_{j},z_{j},v_{j}\right) ,  \label{17}
\end{equation}%
which is in general not linear or separable in $v_{j}$. Working off this
normalization, and essentially repeating the same steps as above holding $%
z_{j}$ fixed, lead to the conclusion that for each $z_{j}$,%
\begin{equation}
-\frac{\partial \omega _{j}\left( a_{j},a_{0},z_{j}\right) }{\partial a_{0}}/%
\frac{\partial \omega _{j}\left( a_{j},a_{0},z_{j}\right) }{\partial a_{j}}%
\equiv \frac{\partial }{\partial a_{0}}q_{j}\left( \mathbf{a,z}\right) /%
\frac{\partial }{\partial a_{j}}q_{0}\left( \mathbf{a,z}\right) \text{.}
\label{111}
\end{equation}%
The RHS of (\ref{111}) is observable from the data, and for each fixed $%
z_{j} $, is solely a function of $a_{0}$, $a_{j}$, which is a testable
implication. If this implication is not rejected, denote the RHS of (\ref%
{111}) as $t_{j}\left( a_{j},a_{0},z_{j}\right) $, just as above. Then for
each each fixed $z_{j}$, solve the PDE%
\begin{equation*}
\frac{\partial \omega _{j}\left( a_{j},a_{0},z_{j}\right) }{\partial a_{0}}+%
\frac{\partial \omega _{j}\left( a_{j},a_{0},z_{j}\right) }{\partial a_{j}}%
t_{j}\left( a_{j},a_{0},z_{j}\right) =0\text{,}
\end{equation*}%
to obtain the $\omega _{j}\left( a_{j},a_{0},z_{j}\right) $, invert w.r.t. $%
a_{0}$ to obtain the utilities $w_{j}\left( a_{j},v_{j},z_{j}\right) $ and
the joint density of $\mathbf{v}$ using the analog of (\ref{5}), where we
utilize the inverse of $\omega _{j}\left( a_{j},a_{0},z_{j}\right) $ w.r.t. $%
a_{j}$, analogous to (\ref{7'}).\footnote{%
If even \textit{conditional} on covariates, independence of preferences and
budget sets issuspect, then one needs to employ a \textquotedblleft control
function\textquotedblright\ type strategy (c.f. Blundell and Powell, 2004)
to estimate the structural choice-probabilities. Indeed, our results above
explore the connection between random utility models and \textquotedblleft
structural\textquotedblright\ choice probabilities. So, given the extensive
econometric literature on estimating structural parameters under
endogeneity, we refrain from discussing the consistent estimation of $%
q_{j}\left( \mathbf{\cdot }\right) $ any further.}

\subsection{Empirical Implications: Bounds on Counterfactuals}

A key empirical implication of our results is that they can be used to
obtain bounds for predicted demand on counterfactual budget sets. We
demonstrate how to construct such bounds in the two leading cases of
interest, viz. price changes and elimination/addition of alternatives.

\textbf{Price Changes}: Denote the support of observed price and income by $%
\mathcal{A}$ and suppose we have to \textit{predict }demand for alternative
1 at a counterfactual $\mathbf{a}^{\prime }=\left( a_{0}^{\prime
},a_{1}^{\prime }...,a_{J}^{\prime }\right) \notin \mathcal{A}$. Such
counterfactual budget sets may arise due to potential price changes, e.g.
those caused by taxes and subsidies or firm-mergers (c.f. Berry and Pakes
1993). To predict this counterfactual demand, let $\mathcal{A}_{j}$ denote
the set of values of $a_{j}$'s that appear in $\mathcal{A}$, and $\mathcal{A}%
_{jk}$ denote the collection of values taken by the pairs $\left\{
a_{j},a_{k}\right\} $, $j\neq k$ that appear in $\mathcal{A}$. Now, using
Lemma 1, we obtain the utility functions $w_{j}\left( a_{j},v_{j}\right) $, $%
j=1,2,...,J$ for $a_{j}\in \mathcal{A}_{j}$, and the joint distribution $%
f\left( \cdot \right) $ of the unobserved heterogeneity $\left\{
v_{1},v_{2},...,v_{j}\right\} $. Recall that our parameter of interest is%
\begin{eqnarray*}
&&q_{1}\left( \mathbf{a}^{\prime }\right) \\
&=&\int 1\left\{ w_{1}\left( a_{1}^{\prime },v_{1}\right) \geq a_{0}^{\prime
},w_{1}\left( a_{1}^{\prime },v_{1}\right) \geq w_{2}\left( a_{2}^{\prime
},v_{2}\right) ,...,w_{1}\left( a_{1}^{\prime },v_{1}\right) \geq
w_{J}\left( a_{J}^{\prime },v_{J}\right) \right\} f\left( \mathbf{v}\right) d%
\mathbf{v}\text{.}
\end{eqnarray*}%
Now, for the pair $\left( a_{1},a_{2}\right) \in \mathcal{A}_{12}$, we have
that $w_{1}\left( a_{1},v_{1}\right) \geq w_{2}\left( a_{2},v_{2}\right)
\Longrightarrow w_{1}\left( a_{1}^{\prime },v_{1}\right) \geq w_{2}\left(
a_{2}^{\prime },v_{2}\right) $ whenever $a_{1}^{\prime }\geq a_{1}$, $%
a_{2}\geq a_{2}^{\prime }$ for any pair $\left( a_{1},a_{2}\right) $.
Accordingly, define $w_{0}\left( a_{0}^{\prime },v_{0}\right) \equiv
a_{0}^{\prime }$, and for each $j=0,2,...,J$ and the upper and lower bound
for $1\left\{ w_{1}\left( a_{1}^{\prime },v_{1}\right) \geq w_{j}\left(
a_{j}^{\prime },v_{j}\right) \right\} $ by%
\begin{eqnarray}
l\left( a_{1}^{\prime },a_{j}^{\prime },v_{1},v_{j}\right) &=&\sup_{ 
_{\substack{ \left( a_{1},a_{j}\right) \in \mathcal{A}_{1j}  \\ a_{1}\leq
a_{1}^{\prime },a_{j}^{\prime }\geq a_{j}}}}1\left\{ w_{1}\left(
a_{1},v_{1}\right) \geq w_{j}\left( a_{j},v_{j}\right) \right\}  \notag \\
u\left( a_{1}^{\prime },a_{j}^{\prime },v_{1},v_{j}\right) &=&\inf 
_{\substack{ \left( a_{1},a_{j}\right) \in \mathcal{A}_{1j}  \\ a_{1}\geq
a_{1}^{\prime },a_{j}^{\prime }\geq a_{j}}}1\left\{ w_{1}\left(
a_{1},v_{1}\right) \geq w_{j}\left( a_{j},v_{j}\right) \right\} \text{.}
\label{20}
\end{eqnarray}%
Therefore, lower and upper bounds on $q_{1}\left( \mathbf{a}^{\prime
}\right) $ are given by%
\begin{eqnarray}
LB_{1}\left( \mathbf{a}^{\prime }\right) &=&\int \left[ \prod\limits 
_{\substack{ j=0  \\ j\neq 1}}^{J}l\left( a_{1}^{\prime },a_{j}^{\prime
},v_{1},v_{j}\right) \right] f\left( \mathbf{v}\right) d\mathbf{v}  \notag \\
UB_{1}\left( \mathbf{a}^{\prime }\right) &=&\int \left[ \prod\limits 
_{\substack{ j=0  \\ j\neq 1}}^{J}u\left( a_{1}^{\prime },a_{j}^{\prime
},v_{1},v_{j}\right) \right] f\left( \mathbf{v}\right) d\mathbf{v}\text{.}
\label{21}
\end{eqnarray}%
Since the utility functions $w_{j}\left( a_{j},v_{j}\right) $, $j=1,2,...,J$
for $a_{j}\in \mathcal{A}_{j}$, and the joint distribution $f\left( \cdot
\right) $ of the unobserved heterogeneity $\left\{
v_{1},v_{2},...,v_{j}\right\} $ are identified using Lemma 1, so are $%
LB_{1}\left( \mathbf{a}^{\prime }\right) $ and $UB_{1}\left( \mathbf{a}%
^{\prime }\right) $.

To get simultaneous bounds on $\left\{ q_{j}\left( \mathbf{a}^{\prime
}\right) \right\} $, $j=0,...,J$, we have to impose the constraint that the
sum of lower bounds and the sum of upper bounds over $j=0,1,...J$ must equal
1. This amounts to finding the set of $\tilde{q}_{j}\left( \mathbf{a}%
^{\prime }\right) $, $j=0,...,J$ such that%
\begin{equation}
\begin{array}{c}
LB_{j}\left( \mathbf{a}^{\prime }\right) \leq \tilde{q}_{j}\left( \mathbf{a}%
^{\prime }\right) \leq UB_{j}\left( \mathbf{a}^{\prime }\right) \text{, }%
\sum\limits_{j=0}^{J}\tilde{q}_{j}\left( \mathbf{a}^{\prime }\right) =1\text{%
,}%
\end{array}
\label{23}
\end{equation}%
where $LB_{j}\left( \mathbf{a}^{\prime }\right) $ and $UB_{j}\left( \mathbf{a%
}^{\prime }\right) $, defined in (\ref{21}), are point-identified and
satisfy the shape restrictions of Lemma 1 (i). Note that (\ref{23}) is a set
of linear equality/inequality constraints in $\tilde{q}_{j}\left( \mathbf{a}%
^{\prime }\right) $ and can be computed using simplex methods. Molinari 2020
discusses several substantive econometric problems that have such linear
structure. The bounds (\ref{23}) on demand in turn provide bounds for
welfare calculations corresponding to changes in prices or quality of the
products, or addition and elimination of options, since welfare expressions
for such cases are known functionals of choice probabilities, c.f.
Bhattacharya 2018. The bounds in (\ref{23}) are sharp because the choice
probabilities $\left\{ q_{j}\left( \mathbf{a}\right) \cup \tilde{q}%
_{j}\left( \mathbf{a}^{\prime }\right) \right\} _{j=0,...,J}$ on $\mathcal{%
A\cup }\left\{ \mathbf{a}^{\prime }\right\} $ where $\tilde{q}_{j}\left( 
\mathbf{a}^{\prime }\right) $ satisfies (\ref{23}), satisfy all conditions
of Lemma 1 and can therefore be rationalized by the same utility functions
and heterogeneity distribution as those that rationalize $\left\{
q_{j}\left( \mathbf{a}\right) \right\} _{j=0,...,J}$ on $\mathcal{A}$.

Allen and Rehbeck 2019 derive bounds for $\left\{ q_{j}\left( \mathbf{a}%
^{\prime }\right) \right\} _{j=0,...,J}$ when $\mathbf{a}^{\prime }\notin 
\mathcal{A}$ by assuming the additive structure $w_{j}\left(
a_{j},v_{j}\right) =w_{j}\left( a_{j}\right) +v_{j}$ and that $w_{j}\left(
a_{j}^{\prime }\right) $ is known even if $\mathbf{a}^{\prime }\notin 
\mathcal{A}$. This is possible if $w_{j}\left( \cdot \right) $ and the joint
distribution of unobserved heterogeneity are parametrically specified, and
the values of these parameters are known from the observed choice
probabilities. In contrast, the bounds in (\ref{23}) do not require such
arbitrary parametric restrictions on the utility indices $w_{j}\left( \cdot
,\cdot \right) $.

\textbf{Change in Choice Sets}: From an initial situation described by the
set-up, suppose alternative $J$ is eliminated from the choice-set. Then the
choice probability $q_{j}\left( \mathbf{a\backslash }\left\{ J\right\}
\right) $ of alternative $j\in \left\{ 0,1,2,...,J-1\right\} $ can be
obtained as follows. First the utilities $w_{j}\left( a_{j},v_{j}\right) $
and the joint density $f_{\mathbf{v}}\left( v_{1},..v_{J-1},v_{J}\right) $
are obtained by applying Lemma 1 to the original choice probabilities when
the entire choice set was available. Then the joint density $f_{\mathbf{v}%
_{-J}}\left( v_{1},..v_{J-1}\right) $ is obtained as%
\begin{equation*}
\int_{-\infty }^{\infty }f_{\mathbf{v}}\left(
v_{1},v_{2},...,v_{J-1},v_{J}\right) dv_{J}
\end{equation*}%
Finally, the choice probability $q_{j}\left( \mathbf{a\backslash }\left\{
J\right\} \right) $ of alternative $j\in \left\{ 0,1,2,...,J-1\right\} $ is
obtained as%
\begin{equation}
\int_{-\infty }^{\infty }...\int_{-\infty }^{\infty }\prod\limits_{k=0,k\neq
j}^{J-1}1\left\{ w_{j}\left( a_{j},v_{j}\right) \geq w_{k}\left(
a_{k},v_{k}\right) \right\} f_{\mathbf{v}_{-J}}\left( v_{1},..v_{J-1}\right)
dv_{J-1}...dv_{1}\text{,}  \label{22}
\end{equation}%
which is point-identified.

\subsection{Random Coefficient Models}

A popular specification of choice probabilities in applied work is the
`random coefficient' model such as the mixed logit or BLP (c.f. Berry 1994.
McFadden and Train 2000, Gautier and Kitamura 2013). McFadden and Train 2000
show that essentially all choice probability functions generated via utility
maximization can be approximated arbitrarily well by an appropriately
defined mixed multinomial logit model. In a random coefficient setting, the
utility of the $i$th individual from choosing the $j$th alternative is
specified as%
\begin{equation*}
U_{ij}=\eta _{ij0}+\sum_{k=1}^{K}z_{jk}\eta _{ik}+\eta _{ip}U\left(
y_{i}-p_{j}\right) \text{,}
\end{equation*}%
where $\mathbf{z}_{j}=\left\{ z_{j1},...,z_{jK}\right\} _{j=1,...,J}$
represents a vector of $K$ observed characteristics of alternative $j$, and $%
\left( \eta _{ij0},\eta _{i1},...,\eta _{iK},\eta _{ip}\right) $ is a random
coefficient vector where $\eta _{ip}>0$ with probability 1 (reflecting
non-satiation in the quantity of numeraire), and $U\left( \cdot \right) $ is
a potentially nonlinear, unknown sub-utility function.\footnote{%
If $U\left( \cdot \right) $ is linear, then income drops out of choice
probabilities, which is a strong and testable restriction.} Then%
\begin{eqnarray*}
U_{ij} &>&U_{il} \\
&\Leftrightarrow &\eta _{ij0}+\sum_{k=1}^{K}z_{jk}\eta _{ik}+\eta
_{ip}U\left( y_{i}-p_{j}\right) >\eta _{il0}+\sum_{k=1}^{K}z_{lk}\eta
_{ik}+\eta _{ip}U\left( y_{i}-p_{l}\right)  \\
&\Leftrightarrow &U\left( y_{i}-p_{j}\right) +\underset{\varepsilon
_{ji}\left( \mathbf{z}_{j}\right) }{\underbrace{\frac{\eta _{ij0}}{\eta _{ip}%
}+\sum_{k=1}^{K}z_{jk}\frac{\eta _{ik}}{\eta _{ip}}}}>U\left(
y_{i}-p_{l}\right) +\underset{\varepsilon _{li}\left( \mathbf{z}_{l}\right) }%
{\underbrace{\frac{\eta _{il0}}{\eta _{ip}}+\sum_{k=1}^{K}z_{lk}\frac{\eta
_{ik}}{\eta _{ip}}}}
\end{eqnarray*}%
which amounts to choice based on the utility functions $U\left(
y_{i}-p_{j}\right) +\varepsilon _{ji}\left( \mathbf{z}_{j}\right) $.
Therefore, for each realization of $\mathbf{z}=\left\{ \mathbf{z}%
_{j}\right\} _{j=1,...,J}$, the conditions and thus conclusions of Theorem 1
hold; the only difference is that the structural choice probabilities
appearing in the statement of the theorem will have to be defined
conditional on $\mathbf{z}=\left\{ \mathbf{z}_{j}\right\} _{j=1,...,J}$.
Similarly, the identification argument of Sec 4.1 will work conditional on $%
\mathbf{z}$, implying that the joint distribution of $\left\{ \varepsilon
_{ji}\left( \mathbf{z}_{j}\right) \right\} $, $j=1,...,J$ is exactly
identified while $U\left( \cdot \right) $ may be over-identified. For
example, one would expect the characteristics of alternatives viz. $\mathbf{z%
}$ to remain identical across consumers in a single market (e.g. the
frequency of various modes of public transport are likely to be identical
across individuals in the same locality). Then the $q_{j}\left( \mathbf{%
\cdot ;z}\right) $'s and their partial derivatives are identified via the 
\emph{variation in income }$y$, and hence in $a_{0}=y$ and $a_{j}\equiv
y-p_{j}$ for $j=1,...,J$, across individuals in the same market and,
additionally, any variation in price within and across markets with the same
observed $\mathbf{z}$'s. Applying the identification argument outlined in
Section \ref{Id}, one obtains the distribution of $\left\{ \varepsilon
_{ji}\left( \mathbf{z}_{j}\right) \right\} $, $j=1,...,J$ conditional on
each realization of $\mathbf{z}$ and the utility indices. These objects will
yield bounds on choice probabilities when the budget set takes
counterfactual values due to potential potential policy interventions, by
applying (\ref{21}) or (\ref{23}) conditional on the $\mathbf{z}^{\prime }$s.

Note further that knowledge of the distribution of the (suitably normalized) 
$\eta ^{\prime }$s will allow one to bound choice probabilities when\textit{%
\ not only the budget set but also covariates take counterfactual values}.
If the number of markets is large, the distribution of random coefficients
is identical in each market, and there is sufficient independent variation
of the $\mathbf{z}$'s across markets, then one can identify the distribution
of the normalized $\eta $s from the distribution of the $\varepsilon
_{j}\left( \mathbf{z}_{j}\right) $'s by using the Cramer-Wold theorem (c.f.
Billingsley 1995, Theorem 29.4, Beran and Hall 1992). To see this, let the
value of $\mathbf{z}_{j}$ in market $m$ be denoted by $\mathbf{z}_{j}^{m}$,
and denote $\varepsilon _{ji}\left( \mathbf{z}_{j}\right) =\gamma
_{i}^{\prime }\mathbf{z}_{j}^{m}$, where $\mathbf{z}_{j}^{m}$ is observed,
and the object of interest is the distribution of the unobserved random
coefficients $\gamma $ which are the normalized values of the $\eta $'s.
Then, using Lemma 1, we obtain the joint distribution of $\left( \gamma
^{\prime }\mathbf{z}_{1}^{m},...,\gamma ^{\prime }\mathbf{z}_{J}^{m}\right) $
in market $m$, and therefore the marginal of $\gamma ^{\prime }\mathbf{z}%
_{1}^{m}$. Doing this in each market gives us the marginal distribution of
each of the projections $\left\{ \gamma ^{\prime }\mathbf{z}_{1}^{m}\right\} 
$, $m=1,...,M$. Now applying the approach of Beran and Hall 1992 as $%
M\rightarrow \infty $ identifies the distribution of $\gamma $ under
appropriate regularity conditions. The precision of the corresponding
estimator can be increased by using information on all $J$ alternatives,
i.e. $\left\{ \gamma ^{\prime }\mathbf{z}_{j}^{m}\right\} ,m=1,...,M$, $%
j=1,2,...,J$.

We conclude this subsection with the observation that Lemma 1 also applies
to more general models e.g. where utilities are given by%
\begin{equation}
U_{ij}=\eta _{ip}U\left( y_{i}-p_{j},\mathbf{z}_{j}\right) +\varepsilon
_{ji}\left( \mathbf{z}_{j}\right) \text{,}
\end{equation}%
where $\eta _{ip}>0$ with probability 1, and the unobserved $\varepsilon
_{ji}\left( \mathbf{z}_{j}\right) $ is not necessarily linear in $\mathbf{z}%
_{j}$. Condition (ii) of Lemma 1, conditional on observed covariates, is
therefore a testable implication of all such models.

\section{Conclusion}

This paper provides a unified analysis of integrability and identification
in multinomial discrete choice models. It establishes closed-form
shape-restrictions on choice-probability functions, under which multinomial
choice probabilities can be rationalized via random utility models. These
conditions are shown to be necessary and sufficient for the additive random
utility model of McFadden. Our results apply equally to random coefficient
models like mixed logit -- widely used in IO applications -- because
conditional on observed characteristics, these are observationally
equivalent to models with additive heterogeneity. Our theoretical results
are obtained via application of the classical theory of partial differential
equations, whose use in economics and econometrics is relatively novel. The
key empirical implications of our results are that they lead to (a)
nonparametric identification of random utility models using economic theory
as opposed to statistical assumptions, (b) specification of multinomial
choice models in applied work that is consistent with economic theory while
allowing for fully nonparametric utility functions, unobserved heterogeneity
and income-effects, and (c) calculation of theory-consistent nonparametric
bounds for demand and welfare on counterfactual budget sets, e.g. those
arising from price change due to a tax or subsidy, firm-mergers and changes
in the number of available alternatives.\pagebreak 

\begin{center}
{\Large References}
\end{center}

\begin{enumerate}
\item Allen, R. and Rehbeck, J., 2019. Identification with additively
separable heterogeneity. Econometrica, 87(3), pp.1021-1054.

\item Armstrong, M. and Vickers, J., 2015. Which demand systems can be
generated by discrete choice?. Journal of Economic Theory, 158, pp.293-307.

\item Beran, R. and Hall, P., 1992. Estimating coefficient distributions in
random coefficient regressions. The Annals of Statistics, 20(4),
pp.1970-1984.

\item Berry, S. 1994. Estimating discrete-choice models of product
differentiation. The RAND Journal of Economics, pp. 242-262.

\item Berry, S. and Pakes, A. 1993. Some applications and limitations of
recent advances in empirical industrial organization: Merger analysis. The
American Economic Review, 83(2), pp.247-252.

\item Bhattacharya, D., 2018. Empirical welfare analysis for discrete
choice: Some general results. Quantitative Economics, 9(2), pp.571-615.

\item Bhattacharya, D., 2021. The empirical content of binary choice models.
Econometrica, 89(1), pp.457-474.

\item Billingsley, P., 2008. Probability and measure. John Wiley \& Sons.

\item Blundell, R., and James L. Powell (2003): Endogeneity in nonparametric
and semiparametric regression models. Econometric society monographs 36,
312-357.

\item Blundell, R.W. and Powell, J.L. (2004): Endogeneity in semiparametric
binary response models. The Review of Economic Studies, 71(3), 655-679.

\item Coddington EA. 1961. An introduction to ordinary differential
equations. Dover Publishing, New York.

\item Courant, R. 1962. Methods of Mathematical Physics, Vol. 2,
Interscience, New York.

\item Daly, A., and Zachary, S.,1978. Improved multiple choice models. In
Hensher,D., Dalvi,Q.(Eds.), Identifying and Measuring the Determinants of
Mode Choice, Teakfields, London.

\item Dette, H., Hoderlein, S. and Neumeyer, N. 2016. Testing multivariate
economic restrictions using quantiles: the example of Slutsky negative
semidefiniteness. Journal of Econometrics 191(1), pp.129-144.

\item Gentzkow, M., 2007. Valuing new goods in a model with complementarity:
Online newspapers. American Economic Review, 97(3), pp.713-744.

\item Gautier, E. and Kitamura, Y., 2013. Nonparametric estimation in random
coefficients binary choice models. Econometrica, 81(2), pp.581-607.

\item Hausman, J.A. and Newey, W.K. 2016. Individual heterogeneity and
average welfare. Econometrica, 84(3), pp.1225-1248.

\item Kitamura, Y. and Stoye, J. (2016): Nonparametric analysis of random
utility models, Econometrica, 86(6), 1883-1909.

\item Lewbel, A., 2001. Demand Systems with and without Errors. American
Economic Review, 91(3), pp.611-618.

\item Mas-Colell, A., Whinston, M.D. and Green, J.R., 1995. Microeconomic
theory. New York: Oxford university press.

\item Matzkin R., 1993. Nonparametric identification and estimation of
polychotomous choice models. Journal of Econometrics, 58(1-2):137-68.

\item Matzkin, R.L., 2007. Nonparametric identification. Handbook of
econometrics, 6, pp.5307-5368.

\item McFadden, D., 1973. Conditional logit analysis of qualitative choice
behavior.

\item McFadden, D., 1978. Modeling the choice of residential location.
Transportation Research Record, (673).

\item McFadden, D., 1981. Econometric models of probabilistic choice.
Structural analysis of discrete data with econometric applications, 198272.

\item McFadden, D. and Richter, M.K. (1990): Stochastic rationality and
revealed stochastic preference. Preferences, Uncertainty, and Optimality,
Essays in Honor of Leo Hurwicz, Westview Press, 161-186.

\item McFadden, D. and Train, K., 2000. Mixed MNL models for discrete
response. Journal of applied Econometrics, 15(5), pp.447-470.

\item Molinari, F. 2020. Microeconometrics with partial identification,
Handbook of Econometrics, Volume 7, Part A, 2020, Pages 355-486.

\item Zachmanoglou, E.C. and Thoe, D.W., 1986. Introduction to partial
differential equations with applications. Courier Corporation.\pagebreak
\end{enumerate}

\section{Appendix}

Two basic results from the theory of partial and ordinary differential
equations are used to prove Lemma 1; here we state those results. We will
use the notation $C^{1}$ to indicate a function that is once continuously
differentiable.\medskip

\textbf{Result 1 (Method of Characteristics)}: Consider the linear
homogeneous PDE%
\begin{equation}
\frac{\partial \sigma \left( x,y,z\right) }{\partial x}+g_{2}\left(
x,y\right) \frac{\partial \sigma \left( x,y,z\right) }{\partial y}%
+g_{3}\left( x,z\right) \frac{\partial \sigma \left( x,y,z\right) }{\partial
z}=0\text{.}  \label{9}
\end{equation}%
Suppose $g_{2}$ and $g_{3}$ are $C^{1}$ and do not vanish simultaneously.
Then a general solution to this equation is given by%
\begin{equation}
\sigma \left( x,y,z\right) =\phi \left( h_{2}\left( x,y\right) ,h_{3}\left(
x,z\right) \right) \text{,}  \label{19}
\end{equation}%
where $\phi \left( \cdot \right) $ is \textit{any} arbitrary $C^{1}$
function, and $h_{2}\left( x,y\right) =c_{2}$ and $h_{3}\left( x,z\right)
=c_{3}$ are general solutions to the ordinary differential equations%
\begin{equation}
\frac{dx}{1}=\frac{dy}{g_{2}\left( x,y\right) }=\frac{dz}{g_{3}\left(
x,z\right) }\text{,}  \label{8}
\end{equation}%
i.e. $\frac{dy}{dx}=g_{2}\left( x,y\right) $, $\frac{dz}{dx}=g_{3}\left(
x,z\right) $. The ODE (\ref{8}) are known as the "characteristic equations"
of the linear PDE (\ref{9}), and existence of a solution to the PDE (\ref{9}%
) amounts to existence of a solution of the ODE (\ref{8}), c.f. Courant,
1962, Chapter I.5, II.2. The intuitive reason for this is that (\ref{9})
means that the vector $\left( 1,g_{2}\left( x,y\right) ,g_{3}\left(
x,z\right) \right) $ is a tangent to any level curve $\sigma \left(
x,y,z\right) =c$. Therefore, for any parametrization $\left( x\left(
t\right) ,y\left( t\right) ,z\left( t\right) \right) $ defining the level
curve $\sigma \left( x\left( t\right) ,y\left( t\right) ,z\left( t\right)
\right) =c$, the corresponding tangent vector $\left( \frac{dx}{dt},\frac{dy%
}{dt},\frac{dz}{dt}\right) $ equals the vector $\left( 1,g_{2}\left( x\left(
t\right) ,y\left( t\right) \right) ,g_{3}\left( x\left( t\right) ,z\left(
t\right) \right) \right) $. The formal statement of this result, c.f.
Zachmanoglou and Thoe 1986 Theorem 4.1, is that (a) if $S$ is a level set of
the solution $\sigma \left( x,y,z\right) $ of (\ref{9}), then for every
point of $S$, the solution curve of (\ref{8}) passing through that point
lies entirely on $S$; conversely, (b) if at every point $\left(
x_{0},y_{0},z_{0}\right) $, the solution curve of (\ref{8}) passing through $%
\left( x_{0},y_{0},z_{0}\right) $ lies entirely on the level surface of the
function $\sigma \left( x,y,z\right) $ passing through $\left(
x_{0},y_{0},z_{0}\right) $, then $\sigma \left( x,y,z\right) $ is a solution
to (\ref{9}). Sub-statement (a) is proved by showing that for any solution
curve of (\ref{8}) given by the parametrization $\left( x\left( t\right)
,y\left( t\right) ,z\left( t\right) \right) $, we must have that $\frac{dx}{%
dt}=1$, $\frac{dy}{dt}=g_{2}\left( x,y\right) $, $\frac{dz}{dt}=g_{3}\left(
x,z\right) $; therefore,%
\begin{eqnarray*}
&&\frac{d}{dt}\sigma \left( x\left( t\right) ,y\left( t\right) ,z\left(
t\right) \right) \\
&=&\frac{\partial \sigma \left( x,y,z\right) }{\partial x}\frac{dx}{dt}+%
\frac{\partial \sigma \left( x,y,z\right) }{\partial y}\frac{dy}{dt}+\frac{%
\partial \sigma \left( x,y,z\right) }{\partial z}\frac{dz}{dt} \\
&=&\frac{\partial \sigma \left( x,y,z\right) }{\partial x}+g_{2}\left(
x,y\right) \frac{\partial \sigma \left( x,y,z\right) }{\partial y}%
+g_{3}\left( x,z\right) \frac{\partial \sigma \left( x,y,z\right) }{\partial
z}=0\text{.}
\end{eqnarray*}%
Sub-statement (b) is proved by noting that if the solution curve of (\ref{8}%
) is described by the parametrization $\left( x\left( t\right) ,y\left(
t\right) ,z\left( t\right) \right) $, then the vector $\left( \frac{dx}{dt},%
\frac{dy}{dt},\frac{dz}{dt}\right) $ is tangent to that curve; therefore,
the vector $\left( 1,g_{2}\left( x\left( t\right) ,y\left( t\right) \right)
,g_{3}\left( x\left( t\right) ,z\left( t\right) \right) \right) $ is tangent
to the curve $\left( x\left( t\right) ,y\left( t\right) ,z\left( t\right)
\right) $ and hence to the level surface $S$ of $\sigma \left( x,y,z\right) $
because $\left( x\left( t\right) ,y\left( t\right) ,z\left( t\right) \right) 
$ lies on $S$; therefore, we must have that the gradient of $\sigma \left(
x,y,z\right) $ is orthogonal to $\left( 1,g_{2}\left( x,y\right)
,g_{3}\left( x,z\right) \right) $, i.e. (\ref{9}) holds.

In (\ref{19}), $\phi \left( \cdot ,\cdot \right) $ can be chosen to be
strictly increasing in both arguments. A unique choice of $\phi \left( \cdot
,\cdot \right) $ is pinned down by boundary conditions; in our application,
these amount to equating $\phi \left( h_{2}\left( x,y\right) ,h_{3}\left(
x,z\right) \right) $ to observed choice probability functions.

\medskip

\textbf{Result 2 (Solution of the Characteristic ODE)}: The second result
restates a global version of the Picard-Lindel\"{o}f theorem that
establishes conditions for existence of a solution to a first-order ODE.

\textbf{Picard-Lindel\"{o}f Theorem: }\textit{Suppose that a function }$%
g:R\times R\rightarrow R$\textit{\ is continuous, and on each strip }$%
S_{a}=\left\{ \left( x,y\right) :\left\vert x\right\vert \leq a,\text{ }%
\left\vert y\right\vert <\infty \right\} $\textit{, }$g\left( x,y\right) $%
\textit{\ is Lipschitz in }$y$\textit{. Then the ordinary differential
equation }$n^{\prime }\left( x\right) =g\left( x,n\left( x\right) \right) $%
\textit{, has a general solution }$n\left( \cdot \right) :R\rightarrow R$%
\textit{\ with }$n\left( \cdot \right) $\textit{\ being }$C^{1}$\textit{.
(See, for instance, Coddington, 1961, Theorem 9 and corollary).}

This result is proved by showing that under the assumptions of the lemma,
the map $n\left( \cdot \right) :\rightarrow \int_{x_{0}}^{x}g\left(
s,n\left( s\right) \right) ds$ for any arbitrary $x_{0}$ is a contraction,
thereby ensuring, via the Banach fixed point theorem, the existence of $%
n\left( \cdot \right) $ satisfying%
\begin{equation*}
n\left( x\right) =n\left( x_{0}\right) +\int_{x_{0}}^{x}g\left( s,n\left(
s\right) \right) ds.
\end{equation*}

\begin{center}
\textbf{Proof of Lemma 1}
\end{center}

\begin{proof}[Proof]
WLOG take $m=0$, and use condition (ii) of the Lemma to define%
\begin{equation}
t_{j0}\left( a_{j},a_{0}\right) \equiv \frac{\partial }{\partial a_{0}}%
q_{j}\left( \mathbf{a}\right) /\frac{\partial }{\partial a_{j}}q_{0}\left( 
\mathbf{a}\right) \geq 0\text{.}  \label{s}
\end{equation}%
\qquad

Now, because $\sum_{j=0}^{J}q_{j}\left( \mathbf{a}\right) =1$,
differentiating both sides w.r.t. $a_{0}$ gives%
\begin{equation}
\frac{\partial }{\partial a_{0}}q_{0}\left( \mathbf{a}\right) +\sum_{j=1}^{J}%
\frac{\partial }{\partial a_{0}}q_{j}\left( \mathbf{a}\right) =0\text{.}
\label{1}
\end{equation}%
Substituting (\ref{s}) in (\ref{1}), we get the linear, homogeneous, partial
differential equation in $q_{0}\left( \cdot \right) $: 
\begin{equation}
\frac{\partial }{\partial a_{0}}q_{0}\left( \mathbf{a}\right) +\sum_{j=1}^{J}%
\frac{\partial }{\partial a_{j}}q_{0}\left( \mathbf{a}\right) \times
t_{j0}\left( a_{j},a_{0}\right) =0\text{.}  \label{t}
\end{equation}

This PDE can be solved via the method of characteristics (see Result 1
above), giving the characteristic ordinary differential equations:%
\begin{equation}
\frac{da_{j}}{da_{0}}=t_{j0}\left( a_{j},a_{0}\right) \text{,}  \label{v}
\end{equation}%
for $j=1,...,J$. Using the Picard-Lindel\"{o}f theorem (Result 2 above) and
the principle of solving linear homogeneous PDEs, we obtain the general
solutions of (\ref{v}) given by $\omega _{j}\left( a_{j},a_{0}\right) =cons$%
, where $\omega _{j}\left( a_{j},a_{0}\right) $ is differentiable, strictly
increasing in $a_{0}$ and strictly decreasing in $a_{j}$, and satisfies%
\begin{equation}
\frac{\partial \omega _{j}\left( a_{j},a_{0}\right) }{\partial a_{0}}+\frac{%
\partial \omega _{j}\left( a_{j},a_{0}\right) }{\partial a_{j}}t_{j0}\left(
a_{j},a_{0}\right) =0\text{,}  \label{vi}
\end{equation}%
and also, using (\ref{s})%
\begin{equation}
-\frac{\partial \omega _{j}\left( a_{j},a_{0}\right) }{\partial a_{0}}/\frac{%
\partial \omega _{j}\left( a_{j},a_{0}\right) }{\partial a_{j}}\equiv \frac{%
\partial }{\partial a_{0}}q_{j}\left( \mathbf{a}\right) /\frac{\partial }{%
\partial a_{j}}q_{0}\left( \mathbf{a}\right) \text{.}  \label{27}
\end{equation}%
A general solution $q_{0}\left( \mathbf{a}\right) $ is therefore of the form%
\begin{equation}
q_{0}\left( \mathbf{a}\right) =H_{0}\left( \omega _{1}\left(
a_{1},a_{0}\right) ,\omega _{2}\left( a_{2},a_{0}\right) ,...,\omega
_{J}\left( a_{J},a_{0}\right) \right) \text{,}  \label{4}
\end{equation}%
where $H_{0}\left( \cdot \right) $ can be chosen to be strictly increasing
and $C^{1}$ in each argument, and with continuous $J$th order cross partial
derivatives. Since $q_{0}\left( \mathbf{a}\right) $ is observed, the exact
functional form of $H_{0}\left( \cdot \right) $ is pinned down by (\ref{4}),
for any set of solutions $\omega _{j}\left( \cdot ,\cdot \right) $ to the
ODEs (\ref{v}). This corresponds to the so-called "initial condition" in the
PDE nomenclature. In particular, given any $a_{0}$, the value of $%
H_{0}\left( x_{1},x_{2},...x_{J}\right) $ at any vector $\left(
x_{1},x_{2},...x_{J}\right) $ is given by%
\begin{equation}
H_{0}\left( x_{1},x_{2},...x_{J}\right) =q_{0}\left( a_{0},b_{1}\left(
x_{1},a_{0}\right) ,...b_{J}\left( x_{J},a_{0}\right) \right) \text{,}
\label{28'}
\end{equation}%
where $b_{j}\left( x_{j},a_{0}\right) $ is defined by the solution $b$ to%
\begin{equation}
\omega _{j}\left( b,a_{0}\right) =x_{j}  \label{7'}
\end{equation}%
In this construction, the choice of $a_{0}$ is immaterial. That is, for two
choices $a_{0}\neq a_{0}^{\prime }$,%
\begin{eqnarray}
&&q_{0}\left( a_{0},b_{1}\left( x_{1},a_{0}\right) ,...b_{J}\left(
x_{J},a_{0}\right) \right)  \notag \\
&=&H_{0}\left( \omega _{1}\left( b_{1}\left( x_{1},a_{0}\right)
,a_{0}\right) ,\omega _{2}\left( b_{2}\left( x_{2},a_{0}\right)
,a_{0}\right) ,...,\omega _{J}\left( b_{J}\left( x_{J},a_{0}\right)
,a_{0}\right) \right) \text{ from (\ref{4})}  \notag \\
&=&H_{0}\left( x_{1},x_{2},...x_{J}\right)  \notag \\
&&\overset{\text{from (\ref{28'})}}{=}H_{0}\left( \omega _{1}\left(
b_{1}\left( x_{1},a_{0}^{\prime }\right) ,a_{0}^{\prime }\right) ,\omega
_{2}\left( b_{2}\left( x_{2},a_{0}^{\prime }\right) ,a_{0}^{\prime }\right)
,...,\omega _{J}\left( b_{J}\left( x_{J},a_{0}^{\prime }\right)
,a_{0}^{\prime }\right) \right)  \notag \\
&=&q_{0}\left( a_{0}^{\prime },b_{1}\left( x_{1},a_{0}^{\prime }\right)
,...b_{J}\left( x_{J},a_{0}^{\prime }\right) \right) \text{.}  \label{5}
\end{eqnarray}

Having obtained the $\omega _{j}\left( \cdot ,\cdot \right) $'s from (\ref{v}%
) and (\ref{4}), for each $j=1,...J$, define the function $w_{j}\left(
a_{j},v\right) $ by inversion, i.e.%
\begin{equation}
w_{j}\left( a_{j},v\right) =\left\{ a_{0}:\omega _{j}\left(
a_{j},a_{0}\right) =v\right\} \text{.}  \label{7}
\end{equation}%
Note that by construction, $w_{j}\left( a_{j},v\right) $ is strictly
increasing and continuous in $a_{j}$ for each $v$. The $w_{j}\left( \cdot
.\cdot \right) $'s will play the role of `utilities' in our proof of
integrability. Set $w_{0}\left( a_{0},v_{0}\right) \equiv a_{0}$.

We now show how to construct the distribution of heterogeneity. Let $%
\mathcal{\bar{V}}_{j}$ denote the co-domain of $\omega _{j}\left( \cdot
,\cdot \right) $, and let%
\begin{equation*}
\mathcal{V}_{j}=\mathcal{\bar{V}}_{j}\cap \left\{ \omega _{j}\left(
a_{j},a_{0}\right) :\dprod_{j=1}^{J}\left\{ \frac{\partial }{\partial a_{0}}%
\omega _{j}\left( a_{j},a_{0}\right) \times \frac{\partial }{\partial a_{j}}%
\omega _{j}\left( a_{j},a_{0}\right) \right\} \neq 0\right\} \text{,}
\end{equation*}%
and let $\mathcal{V}\equiv \times _{j=1}^{J}\mathcal{V}_{j}$. Now, given any
vector $\mathbf{v}\equiv \left( v_{1},...,v_{J}\right) \in \mathcal{V}$,
define the cumulative distribution function at $\mathbf{v}$ as%
\begin{equation*}
F\left( v_{1},...,v_{J}\right) =q_{0}\left( a_{0},a_{1},...,a_{J}\right) 
\text{,}
\end{equation*}%
where the vector $\left( a_{0},a_{1},...,a_{J}\right) $ satisfies $%
v_{j}=\omega _{j}\left( a_{j},a_{0}\right) $, for each $j=1,...J$. It
follows from (\ref{4}) and (\ref{5}) that this function is well-defined. The
above CDF implies the density function $f:\mathcal{V}\rightarrow \mathbb{R}%
^{+}$:%
\begin{eqnarray}
&&f\left( v_{1},...,v_{J}\right)  \notag \\
&=&\frac{\frac{\partial ^{J}}{\partial a_{1}...\partial a_{J}}q_{0}\left(
a_{0},a_{1},...,a_{J}\right) |_{v_{j}=\omega _{j}\left( a_{j},a_{0}\right) 
\text{, }j=1,...J}}{\dprod_{j=1}^{J}\frac{\partial }{\partial a_{j}}\omega
_{j}\left( a_{j},a_{0}\right) |_{v_{j}=\omega _{j}\left( a_{j},a_{0}\right) 
\text{, }j=1,...J}}  \label{31'} \\
&=&\frac{\frac{\partial ^{J-1}}{\partial a_{1}...\partial a_{k-1}\partial
a_{k+1}...\partial a_{J}}\frac{\partial }{\partial a_{k}}q_{0}\left(
a_{0},a_{1},...,a_{J}\right) |_{v_{j}=\omega _{j}\left( a_{j},a_{0}\right) 
\text{, }j=1,...J}}{\dprod_{j=1}^{J}\frac{\partial }{\partial a_{j}}\omega
_{j}\left( a_{j},a_{0}\right) |_{v_{j}=\omega _{j}\left( a_{j},a_{0}\right) 
\text{, }j=1,...J}}\text{, for any }k\in \left\{ 1,...,J\right\}  \notag \\
&=&\frac{\frac{\partial ^{J-1}}{\partial a_{1}...\partial a_{k-1}\partial
a_{k+1}...\partial a_{J}}\left[ -\underset{\text{does not depend on }%
a_{1}...a_{k-1},a_{k+1}...a_{J}\text{ }}{\underbrace{\frac{\frac{\partial
\omega _{k}\left( a_{k},a_{0}\right) }{\partial a_{k}}}{\frac{\partial
\omega _{k}\left( a_{k},a_{0}\right) }{\partial a_{0}}}}}\times \frac{%
\partial }{\partial a_{0}}q_{k}\left( a_{0},a_{1},...,a_{J}\right) \right]
|_{v_{j}=\omega _{j}\left( a_{j},a_{0}\right) }}{\dprod_{j=1}^{J}\frac{%
\partial }{\partial a_{j}}\omega _{j}\left( a_{j},a_{0}\right)
|_{v_{j}=\omega _{j}\left( a_{j},a_{0}\right) \text{, }j=1,...J}}\text{,
from (\ref{27})}  \notag \\
&=&-\frac{\frac{\partial ^{J-1}}{\partial a_{1}...\partial a_{k-1}\partial
a_{k+1}...\partial a_{J}}\frac{\partial }{\partial a_{0}}q_{k}\left(
a_{0},a_{1},...,a_{J}\right) |_{v_{j}=\omega _{j}\left( a_{j},a_{0}\right) 
\text{, }j=1,...J}}{\frac{\frac{\partial }{\partial a_{0}}\omega _{k}\left(
a_{k},a_{0}\right) }{\frac{\partial }{\partial a_{k}}\omega _{k}\left(
a_{k},a_{0}\right) }\times \dprod_{j=1}^{J}\frac{\partial }{\partial a_{j}}%
\omega _{j}\left( a_{j},a_{0}\right) |_{v_{j}=\omega _{j}\left(
a_{j},a_{0}\right) \text{, }j=1,...J}}  \notag \\
&=&-\frac{\frac{\partial ^{J-1}}{\partial a_{1}...\partial a_{k-1}\partial
a_{k+1}...\partial a_{J}}\frac{\partial }{\partial a_{0}}q_{k}\left(
a_{0},a_{1},...,a_{J}\right) |_{v_{j}=\omega _{j}\left( a_{j},a_{0}\right) 
\text{, }j=1,...J}}{\frac{\partial }{\partial a_{0}}\omega _{k}\left(
a_{k},a_{0}\right) \times \dprod_{j=1,j\neq k}^{J}\frac{\partial }{\partial
a_{j}}\omega _{j}\left( a_{j},a_{0}\right) |_{v_{j}=\omega _{j}\left(
a_{j},a_{0}\right) \text{, }j=1,...J}}\text{.}  \label{6}
\end{eqnarray}%
Since $\frac{\partial ^{J}}{\partial a_{0}\partial a_{1}...\partial
a_{k-1}\partial a_{k+1}\partial a_{J}}q_{k}\left(
a_{0},a_{1},...a_{J}\right) $ has sign $\left( -1\right) ^{J}$ and $\frac{%
\partial }{\partial a_{j}}\omega _{j}\left( a_{j},a_{0}\right) <0$, and $%
\frac{\partial }{\partial a_{0}}\omega _{j}\left( a_{j},a_{0}\right) >0$ on $%
\mathcal{V}$, each of the above expressions has numerator and denominator of
the same sign, and is thus non-negative. We verify below that this joint
density integrates to 1.

We now show that the above construction of $w_{j}\left( \cdot ,\cdot \right) 
$ (c.f. (\ref{7})) and the joint density of heterogeneity (\ref{5}) and (\ref%
{6}) will indeed produce the original choice probabilities. To see this for
alternative 1, consider the integral%
\begin{eqnarray*}
&&\int_{\mathcal{V}}1\left\{ w_{1}\left( a_{1},v_{1}\right) \geq \max_{k\in
\left\{ 0,2,...J\right\} }w_{k}\left( a_{k},v_{k}\right) \right\} f\left(
v_{1},v_{2},...,v_{1}\right) dv_{1}...dv_{J} \\
&=&\int_{\mathcal{V}}1\left[ v_{1}\geq \omega _{1}\left( a_{1},a_{0}\right)
,\cap _{k\in \left\{ 2,...J\right\} }1\left\{ v_{k}\leq \omega _{k}\left(
a_{k},w_{1}\left( a_{1},v_{1}\right) \right) \right\} \right] f\left(
v_{1},v_{2},...,v_{1}\right) dv_{1}...dv_{J}
\end{eqnarray*}%
Consider the substitution $\left( v_{1},v_{2},...v_{J}\right) \rightarrow
\left( x_{1},x_{2},...x_{J}\right) $ given by $v_{1}=\omega _{1}\left(
a_{1},x_{1}\right) $ (so that $x_{1}=w_{1}\left( a_{1},v_{1}\right) $), and
for $k=2,...,J$, $v_{k}=\omega _{k}\left( x_{k},x_{1}\right) $, which
transforms the above integral to%
\begin{eqnarray}
&&\int_{a_{0}}^{\infty }\int_{a_{2}}^{\infty }...\int_{a_{J}}^{\infty }\left[
\begin{array}{c}
f\left( \omega _{1}\left( a_{1},x_{1}\right) ,\omega _{2}\left(
x_{2},x_{1}\right) ...,\omega _{J}\left( x_{J},x_{1}\right) \right) \\ 
\times \left\vert \frac{\partial \omega _{1}\left( a_{1},x_{1}\right) }{%
\partial x_{1}}\times \dprod_{k=2}^{J}\frac{\partial \omega _{j}\left(
x_{j},x_{1}\right) }{\partial x_{j}}\right\vert%
\end{array}%
\right] dx_{J}...dx_{2}dx_{1}  \notag \\
&=&\int_{a_{0}}^{\infty }\int_{a_{2}}^{\infty }...\int_{a_{J}}^{\infty } 
\left[ 
\begin{array}{c}
f\left( \omega _{1}\left( a_{1},x_{1}\right) ,\omega _{2}\left(
x_{2},x_{1}\right) ...,\omega _{J}\left( x_{J},x_{1}\right) \right) \\ 
\times \left( -1\right) ^{J-1}\times \frac{\partial \omega _{1}\left(
a_{1},x_{1}\right) }{\partial x_{1}}\times \dprod_{k=2}^{J}\frac{\partial
\omega _{j}\left( x_{j},x_{1}\right) }{\partial x_{j}}%
\end{array}%
\right] dx_{J}...dx_{2}dx_{1}  \notag \\
&=&\left( -1\right) ^{J-1}\times \int_{a_{0}}^{\infty }\int_{a_{2}}^{\infty
}...\int_{a_{J}}^{\infty }\left\{ -\frac{\partial ^{J}}{\partial
x_{1}\partial x_{2}...\partial x_{J}}q_{1}\left(
x_{1},a_{1},x_{2},...x_{J}\right) \right\} dx_{J}...dx_{2}dx_{1}\text{, by (%
\ref{6})}  \notag \\
&=&\left( -1\right) ^{J}\times \int_{a_{0}}^{\infty }\int_{a_{2}}^{\infty
}...\int_{a_{J}}^{\infty }\left\{ \frac{\partial ^{J}}{\partial
x_{1}\partial x_{2}...\partial x_{J}}q_{1}\left(
x_{1},a_{1},x_{2},...x_{J}\right) \right\} dx_{J}...dx_{2}dx_{1}  \notag \\
&=&\int_{\infty }^{a_{0}}\int_{\infty }^{a_{2}}...\int_{\infty
}^{a_{J}}\left\{ \frac{\partial ^{J}}{\partial x_{1}\partial
x_{2}...\partial x_{J}}q_{1}\left( x_{1},a_{1},x_{2},...x_{J}\right)
\right\} dx_{J}...dx_{2}dx_{1}  \notag \\
&=&q_{1}\left( a_{0},a_{1},a_{2},...a_{J}\right) \text{.}  \label{13}
\end{eqnarray}

Exactly analogous steps for $j=2,...J$, and using (\ref{6}), lead to the
conclusion that for all $j\geq 1$,%
\begin{eqnarray*}
&&\int 1\left\{ w_{j}\left( a_{j},v_{j}\right) \geq \max_{k\in \left\{
0,1,2,...J\right\} \backslash \left\{ j\right\} }w_{k}\left(
a_{k},v_{k}\right) \right\} f\left( v_{1},v_{2},...,v_{1}\right)
dv_{1}...dv_{J} \\
&=&q_{j}\left( a_{0},a_{1},a_{2},...a_{J}\right) \text{.}
\end{eqnarray*}

Also, note that%
\begin{eqnarray*}
&&\int 1\left\{ a_{0}\geq \max_{k\in \left\{ 1,2,...J\right\} }w_{k}\left(
a_{k},v_{k}\right) \right\} f\left( v_{1},v_{2},...,v_{J}\right)
dv_{1}...dv_{J} \\
&=&\int_{0}^{\omega _{1}\left( a_{1},a_{0}\right) }...\int_{0}^{\omega
_{J}\left( a_{J},a_{0}\right) }f\left( v_{1},v_{2},...,v_{J}\right)
dv_{J}...dv_{1} \\
\text{substitute }v_{j} &\rightarrow &x_{j}\text{ satisfying }v_{j}=\omega
_{j}\left( x_{j},a_{0}\right) \\
&=&\int_{a_{1}}^{\infty }\int_{a_{2}}^{\infty }...\int_{a_{J}}^{\infty
}f\left( \omega _{1}\left( x_{1},a_{0}\right) ,...,\omega _{J}\left(
x_{J},a_{0}\right) \right) \left\vert \frac{\partial \omega _{1}\left(
x_{1},a_{0}\right) }{\partial x_{1}}...\frac{\partial \omega _{J}\left(
x_{J},a_{0}\right) }{\partial x_{J}}\right\vert dx_{J}...dx_{1} \\
&=&\int_{a_{1}}^{\infty }\int_{a_{2}}^{\infty }...\int_{a_{J}}^{\infty
}\left( -1\right) ^{J}\times f\left( \omega _{1}\left( x_{1},a_{0}\right)
,...,\omega _{J}\left( x_{J},a_{0}\right) \right) \frac{\partial \omega
_{1}\left( x_{1},a_{0}\right) }{\partial x_{1}}...\frac{\partial \omega
_{J}\left( x_{J},a_{0}\right) }{\partial x_{J}}dx_{J}...dx_{1} \\
&=&\int_{\infty }^{a_{1}}...\int_{\infty }^{a_{J}}\frac{\partial ^{J}}{%
\partial \alpha _{1}...\partial \alpha _{J}}q_{0}\left( a_{0},\alpha
_{1},...\alpha _{J}\right) |_{\alpha _{1}=x_{1},...\alpha
_{J}=x_{J}}dx_{J}...dx_{1}\text{, by (\ref{31'})} \\
&=&q_{0}\left( a_{0},a_{1},...a_{J}\right) \text{.}
\end{eqnarray*}

Finally, to show that the joint density (\ref{31'}) integrates to 1, use
exactly the same substitution as the one leading to (\ref{13}), and observe
that%
\begin{eqnarray*}
&&\int f\left( v_{1},v_{2},...,v_{J}\right) dv_{1}...dv_{J} \\
&=&\int_{-\infty }^{\infty }\int_{-\infty }^{\infty }...\int_{-\infty
}^{\infty }\left[ 
\begin{array}{c}
f\left( \omega _{1}\left( a_{1},x_{1}\right) ,\omega _{2}\left(
x_{2},x_{1}\right) ...,\omega _{J}\left( x_{J},x_{1}\right) \right) \\ 
\times \left\vert \frac{\partial \omega _{1}\left( a_{1},x_{1}\right) }{%
\partial x_{1}}\times \dprod_{k=2}^{J}\frac{\partial \omega _{j}\left(
x_{j},x_{1}\right) }{\partial x_{j}}\right\vert%
\end{array}%
\right] dx_{2}...dx_{J}dx_{1} \\
&=&\left( -1\right) ^{J}\times \int_{-\infty }^{\infty }\int_{-\infty
}^{\infty }...\int_{-\infty }^{\infty }\left\{ \frac{\partial ^{J}}{\partial
x_{1}\partial x_{2}...\partial x_{J}}q_{1}\left(
x_{1},a_{1},x_{2},...x_{J}\right) \right\} dx_{2}...dx_{J}dx_{1} \\
&=&q_{1}\left( -\infty ,a_{1},-\infty ,...-\infty \right) \\
&=&1\text{, by condition (i) of Lemma 1.}
\end{eqnarray*}

Thus we have shown that a population endowed with our constructed $%
w_{j}\left( \cdot ,v_{j}\right) $ as utilities, together with the joint
density of heterogeneity given by (\ref{5}) would indeed produce the choice
probabilities $\left\{ q_{j}\left( \cdot ,...\cdot \right) \right\} $ for
each $j=0,1,...J$.\smallskip
\end{proof}

\begin{center}
\textbf{Proof of Theorem 1}
\end{center}

\begin{proof}
Necessity is obvious. In particular, condition (ii') is a direct consequence
of equation (\ref{b}).

To prove sufficiency, WLOG take $m=0$, and let $\bar{G}_{j}\left(
a_{j}\right) $ and $\bar{G}_{0}\left( a_{0}\right) $ be the primitive
integrals of $G_{j}\left( a_{j}\right) $ and $G_{0}\left( a_{0}\right) $,
i.e. $\frac{d}{da_{j}}\bar{G}_{j}\left( a_{j}\right) =G_{j}\left(
a_{j}\right) $ and $\frac{d}{da_{0}}\bar{G}_{0}\left( a_{0}\right)
=G_{0}\left( a_{0}\right) $; note that $\bar{G}_{j}\left( a_{j}\right) $ and 
$\bar{G}_{0}\left( a_{0}\right) $ are strictly increasing and continuous
since they have strictly positive derivatives. Then, by exactly analogous
steps that led to (\ref{4}), we have that condition (ii') of Theorem 1, viz. 
$\frac{\partial }{\partial a_{m}}q_{j}\left( \mathbf{a}\right) /\frac{%
\partial }{\partial a_{j}}q_{m}\left( \mathbf{a}\right) =G_{m}\left(
a_{m}\right) /G_{j}\left( a_{j}\right) $ has a general solution of the form%
\begin{equation*}
q_{0}\left( \mathbf{a}\right) =H\left( \bar{G}_{1}\left( a_{1}\right) -\bar{G%
}_{0}\left( a_{0}\right) ,...\bar{G}_{J}\left( a_{J}\right) -\bar{G}%
_{0}\left( a_{0}\right) \right) \text{,}
\end{equation*}%
where $H\left( \cdot \right) $ is an arbitrary smooth function mapping $%
\mathbb{R}^{J}\rightarrow \left[ 0,1\right] $. In particular, we can take $%
H\left( \cdot \right) $ to be nondecreasing in each argument, and we have
that $\bar{G}_{j}\left( \cdot \right) $, $j=0,...,J$ are strictly increasing
and continuous. Following exactly analogous steps to the proof of Lemma 1,
we get that $q_{j}\left( \mathbf{a}\right) $ is rationalized by the utility
functions $w_{0}\left( a_{0},\eta \right) =a_{0}$, $w_{j}\left( a_{j},\eta
\right) =\bar{G}_{0}^{-1}\left( \bar{G}_{j}\left( a_{j}\right) -v_{j}\right) 
$, with the CDF for the joint distribution of the unobserved heterogeneity $%
\eta \equiv \left( v_{1},...,v_{J}\right) $ given by%
\begin{eqnarray}
F_{\eta }\left( v_{1},v_{2},...,v_{J}\right) &=&q_{0}\left( a_{0},\bar{G}%
_{1}^{-1}\left( \bar{G}_{0}\left( a_{0}\right) +v_{1}\right) ,...\bar{G}%
_{J}^{-1}\left( \bar{G}_{0}\left( a_{0}\right) +v_{J}\right) \right)
\label{10} \\
&=&H\left( v_{1},...,v_{J}\right)  \label{16}
\end{eqnarray}%
Just as in (\ref{5}), the choice of $a_{0}$ is immaterial here. Note further
that the above model is observationally equivalent to one where utilities
are given by $W_{0}\left( a_{0},\eta \right) =\bar{G}_{0}\left( a_{0}\right) 
$, $W_{j}\left( a_{j},\eta \right) =\bar{G}_{j}\left( a_{j}\right) -v_{j}$, $%
j=1,...J$, with the joint CDF of $\eta \equiv \left( v_{1},...,v_{J}\right) $
still given by (\ref{16}). This is precisely the ARUM model. That these
distribution implies%
\begin{equation*}
q_{j}\left( \mathbf{a}\right) =\Pr [\cap _{k\neq j}W_{j}\left( a_{j},\eta
\right) \geq W_{k}\left( a_{k},\eta \right) ]
\end{equation*}%
for all $j=0,...,J$ can be established following the exact same steps as in
the proof of Lemma 1 above, with $\omega _{j}\left( a_{j},a_{0}\right) $
replaced by $\bar{G}_{j}\left( a_{j}\right) -\bar{G}_{0}\left( a_{0}\right) $
everywhere.
\end{proof}

\end{document}